\documentclass[pre,twocolumn,floatfix,superscriptaddress,longbibliography,nofootinbib]{revtex4-1}
\RequirePackage[sort&compress]{natbib}

\usepackage{bbold}
\usepackage{bbm}
\usepackage[pdftex]{graphicx}
\usepackage{latexsym,amsmath,verbatim}
\usepackage{color}
\usepackage{rotating}
\usepackage{multirow}
\usepackage[english]{babel}
\usepackage{color}
\usepackage{changes}
\usepackage{hyperref}
\usepackage{booktabs}

\usepackage{lineno}

\usepackage{subfigure}

\usepackage{latexsym,amsmath,verbatim}
\usepackage{mathrsfs}
\usepackage{mathtools}

\usepackage{boxedminipage}

\usepackage{multirow}

\usepackage{ifpdf}

\usepackage{bm}

\newcommand{\heading}[1]{\noindent\textbf{#1.}}

\ifpdf
\usepackage[pdftex]{graphicx}
\else
\usepackage{graphicx}
\fi

\ifpdf 
\DeclareGraphicsExtensions{.pdf,.jpg,.png}
\else
\DeclareGraphicsExtensions{.eps,.jpg}
\fi

\begin{document}

\title{Mapping multiplex hubs in human functional brain network}

\author{Manlio De Domenico$^{1}$, Shuntaro Sasai$^{2}$, Alex Arenas$^{1}$\\
\normalsize{$^{1}$Departament d'Enginyeria Inform\`{a}tica i Matem\`{a}tiques, Universitat Rovira i Virgili, 43007 Tarragona, Spain,\\
$^{2}$Department of Psychiatry, University of Wisconsin - Madison, Madison, WI, USA}
}

\begin{abstract}
Typical brain networks consist of many peripheral regions and a few highly central ones, i.e. hubs, playing key functional roles in cerebral inter-regional interactions. Studies have shown that networks, obtained from the analysis of specific frequency components of brain activity, present peculiar architectures with unique profiles of region centrality. However, the identification of hubs in networks built from different frequency bands simultaneously is still a challenging problem, remaining largely unexplored. Here we identify each frequency component with one layer of a multiplex network and face this challenge by exploiting the recent advances in the analysis of multiplex topologies. First, we show that each frequency band carries unique topological information, fundamental to accurately model brain functional networks. We then demonstrate that hubs in the multiplex network, in general different from those ones obtained after discarding or aggregating the measured signals as usual, provide a more accurate map of brain's most important functional regions, allowing to distinguish between healthy and schizophrenic populations better than conventional network approaches.
\end{abstract}

\maketitle

The brain functional network is generally built by interconnecting brain regions according to some measure of functional connectivity~\cite{bassett2006small,bullmore2009complex,bullmore2012economy}. Studies using functional magnetic resonance imaging~\cite{van2010exploring,poldrack2015progress} (fMRI) provided convincing evidences supporting the existence of special regions, i.e. hubs, that play a fundamental role in brain functional connectivity~\cite{achard2006resilient,power2013evidence} by mediating interactions among other regions and favoring the brain's integrated operation. Generally, the strength of this connectivity is empirically estimated by inter-regional correlations calculated after post-processing and filtering fMRI signals with a conventional pass band, keeping components between 0.01 and 0.1~Hz~\cite{cordes2001frequencies,cordes2002hierarchical,fox2007spontaneous}. The importance of each region with respect to the overall connectivity, i.e. nodal centrality in the functional network, is of particular interest in many applications~\cite{sporns2007identification,bassett2008hierarchical,bullmore2009complex,lynall2010functional,rubinov2010complex,zuo2012network}. However, it has been shown that networks with unique hub regions can be built from different frequency ranges~\cite{sasai2014frequency} and that region centrality might largely fluctuate depending on frequency cuts~\cite{thompson2015frequency}, with components above 0.1~Hz also contributing to functional connectivity with unique topological information~\cite{bassett2006adaptive,mantini2007electrophysiological,supekar2008network,chavez2010functional,liao2013functional,chen2015bold}. Such evidences impel the development of a novel framework to account for full information from all frequency bands separately and simultaneously, without discarding any particular component or aggregating some of them to build single networks. 

In this study, we tackle this challenging issue by employing the theoretical and computational tools recently developed for analyzing and modeling multiplex networks~\cite{mucha2010community,dedomenico2013mathematical,dedomenico2015structural,dedomenico2015ranking}. Multiplex architectures are special networks consisting of different layers, each encoding a different type of relationship or interaction between nodes~\cite{kivela2014multilayer,boccaletti2014structure}. In this context, we identify each frequency component with a distinct layer of a multiplex network whose nodes represent the brain's regions of interest and edges represent their functional connectivity in a specific frequency range. 

This novel approach arises two fundamental questions, requiring to i) verify if and how brain regions playing the role of hubs in the new multiplex functional network differ from the ones obtained using standard network approaches; and ii) if and how we can exploit such differences to improve our understanding of brain disorders. In the following, we will provide extensive evidences demonstrating that hub regions in multiplex functional networks are different from hub regions in standard functional networks and that such differences in the nodal centrality profile allow us to identify patients affected by schizophrenia more accurately than conventional approaches based on discarding or aggregating information about brain functional activity.


\section*{Results}
\heading{Building aggregated and multiplex functional connectivity networks}
We use a publicly available COBRE data set of resting state fMRI, consisting of 71 patients affected by Schizophrenia and 74 healthy controls (age: 18--65). The set of 264 regions of interest (ROIs) introduced by Power \emph{et al.}~\cite{power2011functional} is used to extract the mean signal within each ROI, for each individual separately. After estimating coherence between all pairs of ROIs, the frequency-specific connectivity matrices are obtained by averaging coherence within 12 frequency bands, defined by decomposing the frequency range from 0.01 to 0.25~Hz into intervals with equal widths of 0.02~Hz. The upper bound of this frequency range corresponds to the Nyquist frequency of fMRI signals, while the lower bound is obtained by following conventional way to eliminate long term drift~\cite{cordes2002hierarchical}. 

\begin{figure}[!t]
\centering
\includegraphics[width=9cm]{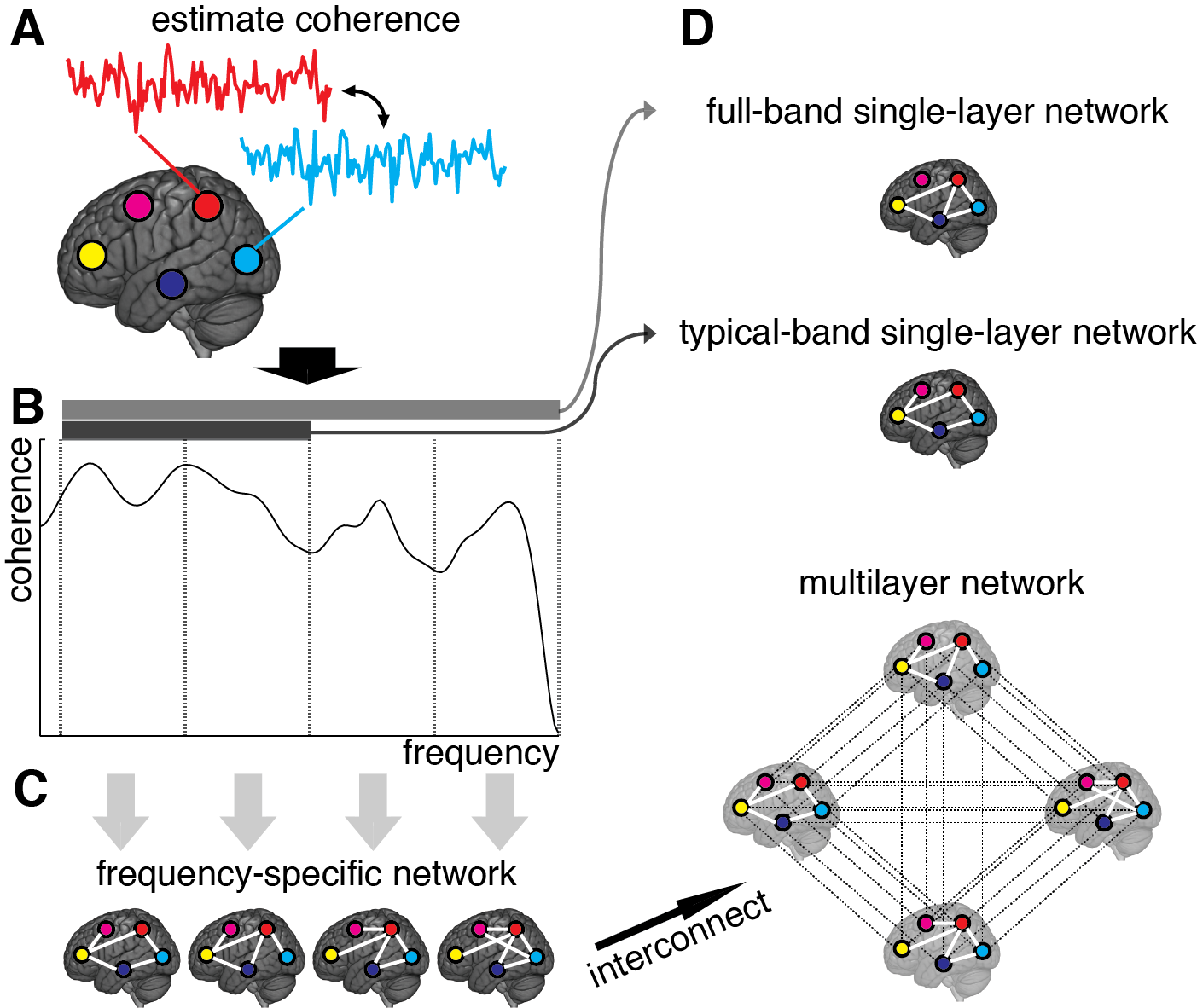}
\caption{\textbf{Schematic illustration of brain multiplex functional network construction.} (A) We measure the brain activity with a set of 264 ROIs (here, we only draw five ROIs, for simplicity), and estimate the coherence spectrum of signals between any pair of ROIs. (B) Averaged coherence values are calculated in 12 frequency bands (here we only show four bands, for simplicity), to quantify the strength of frequency-specific functional connectivity. The statistical significance of each connection is calculated (see Methods) and connections with Z-score smaller than 3 are discarded. (C) The remaining connections are used to build adjacency matrices, weighted by Z-scores, that constitute the layers of the multiplex functional network once interconnected. (D) Resulting single-layer and multiplex networks obtained from this procedure.}\label{fig:scheme}
\end{figure}

Weighted adjacency matrices, defining the functional network for each frequency component separately, are yielded by discarding from frequency-specific connectivity matrices those connections with non-significant amount of correlation (see Methods). The resulting multiplex network is obtained, for each individual separately in control and patient groups, by interconnecting the layers encoding functional connectivity in each frequency band (Fig.~\ref{fig:scheme}A--C). We also define two single-layer networks, obtained by averaging coherence signals within 0.01-0.25~Hz and 0.01-0.1~Hz frequency ranges (Fig.~\ref{fig:scheme}D). We refer to such conventional networks as full-band and typical-band single-layer networks, respectively, both representing averaged and filtered versions of the full multiplex functional networks. 

\begin{figure}[!h]
\centering
\includegraphics[width=9cm]{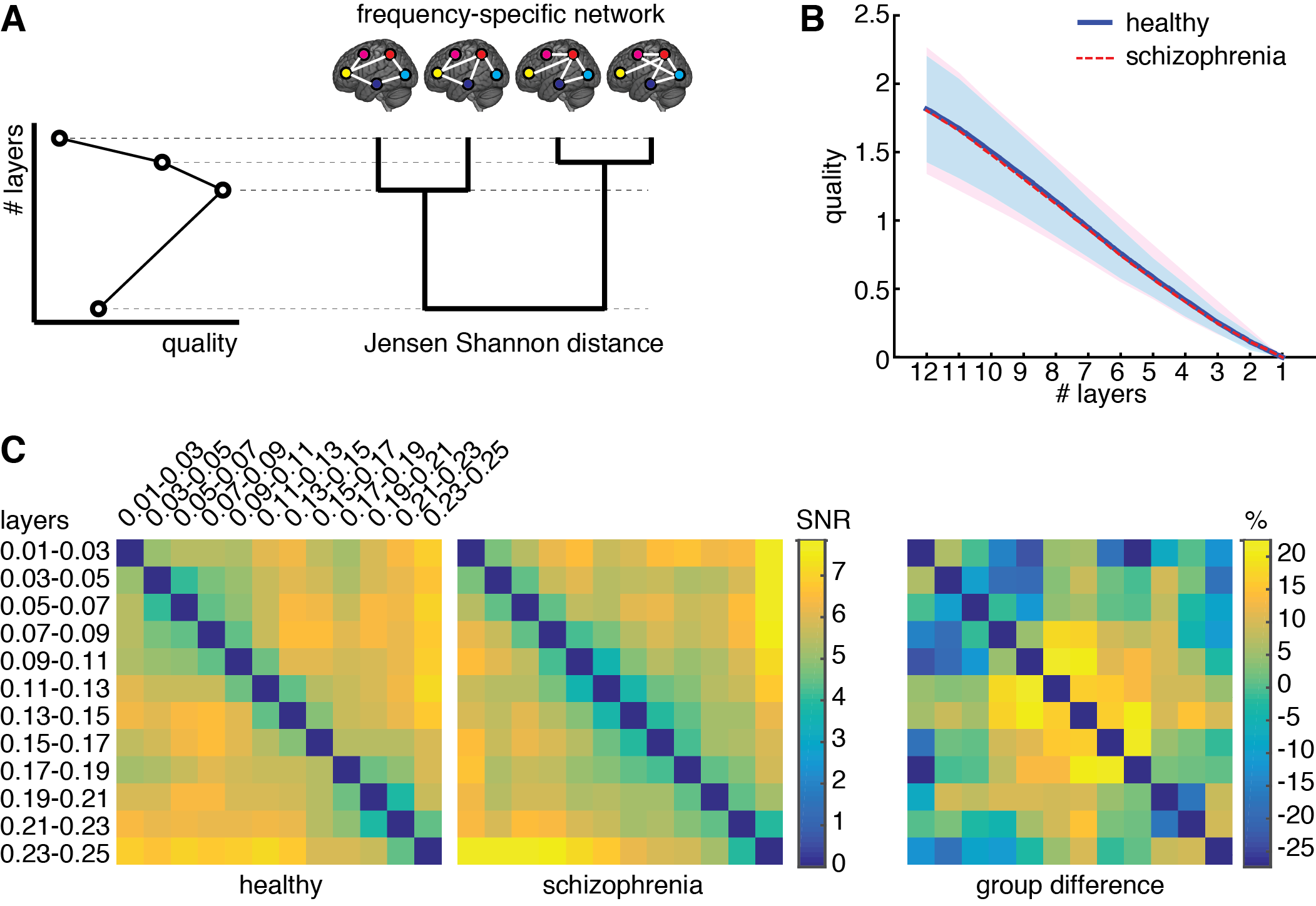}
\caption{\textbf{Structural reducibility of the multiplex functional network.} (A) Schematic illustration of how the analysis structural reducibility of the network works: it allows to identify frequency bands providing redundant topological information and to verify the validity of the multiplex model with respect to conventional single-layer models. Global maxima in the quality function identify optimal structural reductions. (B) The median quality function is shown for healthy control (solid) and schizophrenic patients (dashed), with shaded areas indicating the standard deviation  around each value. (C) Signal-to-noise ratio (SNR; see text for further details) for Jensen-Shannon distance calculated for each pair of layers, color-coded for both groups, and corresponding relative difference between the two groups.}\label{fig:reducibility}
\end{figure}

\heading{Structural reducibility of the multiplex functional connectivity network}
First, we verify if the multiplex network is a valid and suitable model of the underlying brain connectivity. For this purpose, we analyze the structural reducibility of a multiplex network~\cite{dedomenico2015structural}, allowing to identify layers carrying redundant topological information. The method incorporates redundant layers into other ones to reduce the overall structure, while still maximizing the distinguishability between the multiplex network model and the corresponding fully aggregated graph, obtained by summing up the connectivity of all layers (Fig.~\ref{fig:reducibility}A). The difference between connectivity in different layers is quantified by Jensen-Shannon divergence (see Methods), a powerful information-theoretical measure of (dis)similarity. A quality function controls the reduction process and its global maxima identify optimal structural reduction strategies.

We perform structural reducibility for each individual separately and calculate the corresponding quality functions, for control and patients groups (Fig.~\ref{fig:reducibility}B). In both cases, we found that the maximum value of the quality function is attained when no reduction is performed at all, providing evidence that the topological information carried by each functional network, corresponding to a different frequency component, should not be disregarded from structural analyses. It is worth noting that the behavior of the quality function alone does not allow to distinguish between the two groups of individuals.

To gain insights about (dis)similarities between different layers of the multiplex functional network in the two groups, we use the quantum Jensen-Shannon distance (see Methods) calculated during the structural reducibility analysis. The distance matrix, whose entries provide the Jensen-Shannon distance between any pairs of layers, is first built for each individual separately, and group average $\mu$ and standard deviation $\sigma$ are calculated. The signal-to-noise ratio (SNR) defined by their ratio is successively calculated for each pair of layers and for each group, separately (Fig.~\ref{fig:reducibility}C), as well as the relative difference between the two group-averaged values. We observed differences of up to 30\% in absolute value between the two groups, for specific pairs of layers. Dissimilarities between layers within the typical-band were higher in healthy individuals than in schizophrenic patients. On one hand, functional connectivity in healthy subjects is rather volatile and, in general, exhibits topological differences across individuals~\cite{sasai2014frequency} that we did not observe in patients, suggesting the possibility that schizophrenia might alter brain's integrated operation to reduce such a functional diversity. On the other hand, an abnormal amount of dissimilarity between functional networks corresponding to other frequency bands (such as those within relatively higher ranges, e.g. 0.09-0.19~Hz) was observed in patients but not in healthy individuals. These results suggest that the dependence on frequency of patients' functional connectivity is different from that of healthy individuals and we might use such dissimilarity patterns as a fingerprint of brain's functional organization for each group.

\begin{figure}[!h]
\centering
\includegraphics[width=9cm]{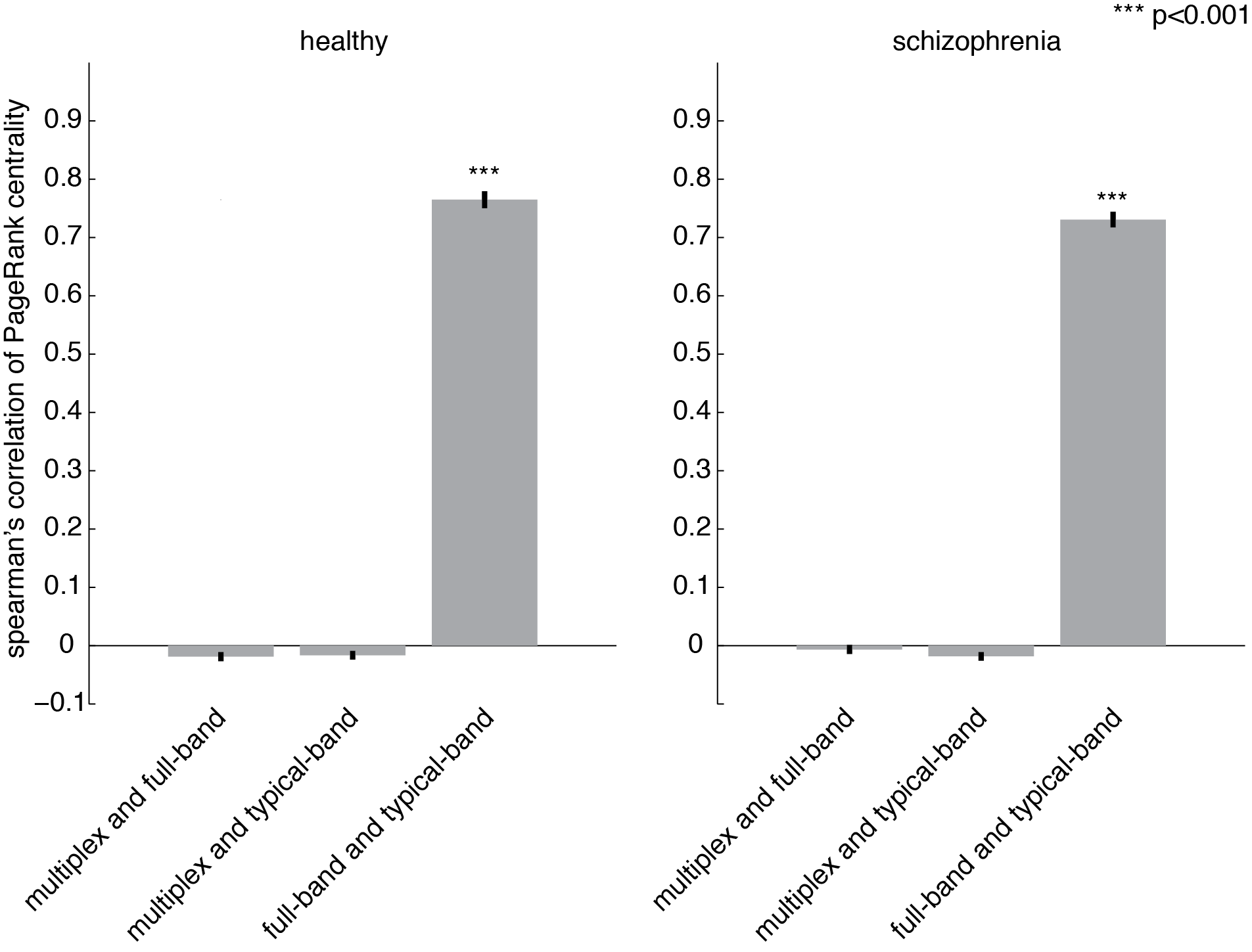}
\caption{\textbf{Comparing centrality profiles of multiplex and conventional functional networks.} Spearman's correlation coefficient between the centrality profiles obtained from multiplex, full-band and typical-band functional networks for (A) healthy and (B) patient groups.}\label{fig:correlation}
\end{figure}

\heading{Identifying schizophrenic patients by the centrality profile of their brain functional connectivity}
The importance of a region with respect to the overall brain functional connectivity can be quantified by centrality descriptors~\cite{achard2006resilient,sporns2007identification,bassett2008hierarchical,lynall2010functional,power2013evidence,van2013network,rubinov2013schizophrenia,thompson2015frequency}. Here, we propose to use PageRank~\cite{brin1998,erman2015google} centrality as a measure of centrality, which is based on the rationale that nodes linked by influential nodes are more central than those linked by un-influential nodes. It has been used in several applications, from ranking relevant Web pages in the World Wide Web~\cite{brin1998} to identifying important nodes in the human functional connectome~\cite{zuo2012network} and, more recently, in a variety of multiplex networks~\cite{dedomenico2015ranking}. Once centrality scores are calculated for each node, the set of all their values constitutes the centrality profile of the underlying functional network. The Spearman's correlation coefficient between the centrality profiles corresponding to the multiplex, full-band and typical-band functional networks is calculated (Fig.~\ref{fig:correlation}A and \ref{fig:correlation}B). While very strong correlations are observed for centrality profiles calculated from single-layer networks, no significant correlation with multiplex centrality profiles were found. 

These results suggest the appealing possibility to use the multiplex centrality profiles to gain new insights about brain functional connectivity. To this aim, we interpret the centrality profiles as characteristic features of each individual (control or patient) and we use the well-known and robust random forest method~\cite{breiman2001random} to train a classifier distinguishing between healthy and schizophrenic individuals (see Methods for further details). At the very beginning, we trained the classifier by using all 264 centrality scores available for each individual and found a classification accuracy of about 60-65\%, regardless for the type of centrality profile used (i.e. multiplex, full-band and typical-band). One of the main advantage of random forest classification is that it also ranks the features based on their classification power, i.e. on the degree of discrimination they have. We capitalize on this precious information to perform a second round of classification, this time using only top-ranked features instead of the full set. We varied between 10 and 50 the number of top features used to discriminate between healthy and schizophrenic individuals. The comparison between the results obtained from different centrality profiles are shown in Fig.~\ref{fig:rforest}A (see Supplementary Figs.~\ref{fig:SI1}--\ref{fig:SI3} for further details). Remarkably, multiplex centrality profiles allow a more accurate discrimination of the two populations, confirming the hypothesis that multiplex functional networks provide a more suitable model of brain functional connectivity. This result is robust against the selection of the number of features used to discriminate, with the multiplex approach significantly outperforming the other ones. 

To gain further insights, we focus on the regions corresponding to the top 30 ROIs of the multiplex centrality profile, where we attained the maximum discrimination between control and patient groups. The spatial distribution of the corresponding brain's regions are shown in Fig.~\ref{fig:rforest}B. Anatomical information about these ROIs is summarized in Supplementary Table~\ref{tab:S1}. 

\heading{Characterizing regions distinctive of schizophrenic brain functional activity}
Capitalizing on results from group discrimination using centrality profiles, we investigate more in detail the role of hub regions. In particular, our interest is twofold. On one hand, we wonder if the most central regions obtained from multiplex and conventional functional networks are the same (it is worth remarking that previous correlation analysis of centrality profiles does not provide this information, because differences might be due to low-ranked regions, for instance). On the other hand we want to clarify which hub regions are found only in healthy individuals, which ones are found only in schizophrenic individuals and which ones are found in both groups.

\begin{figure}[!t]
\centering
\includegraphics[width=9cm]{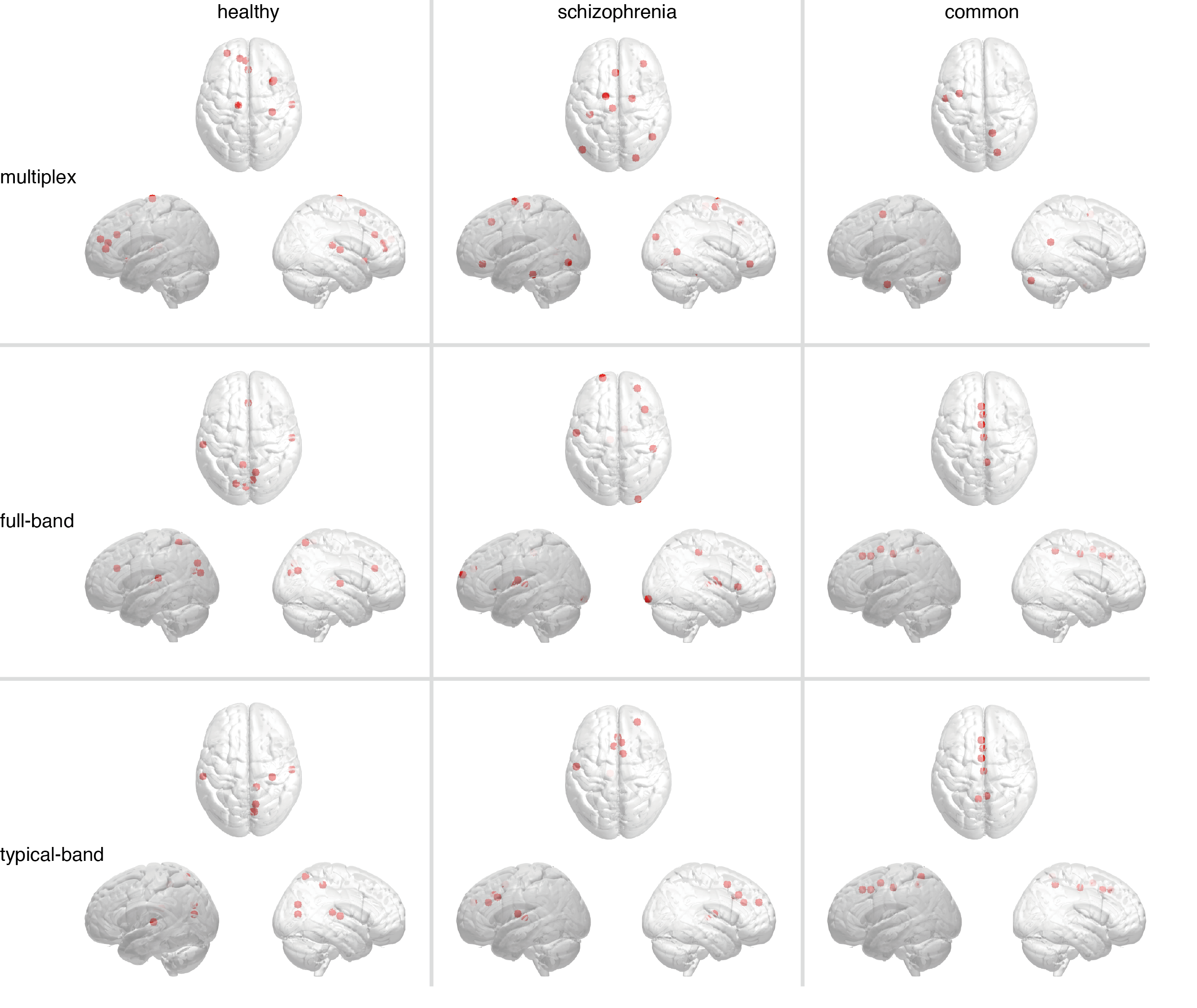}
\caption{\textbf{Brain regions playing the role of hubs in functional connectivity.} The most central regions, i.e. hubs, identified in multiplex and conventional functional networks are shown (from top to bottom). Markers indicate their locations, whereas panels from left-hand to right-hand side show hubs found only in healthy controls (left), only in schizophrenic patients (center), or in both (right).}\label{fig:locations}
\end{figure}

Hubs were identified as ROIs ranked in top 5\% in terms of group-averaged region centrality. Figure~\ref{fig:locations} shows the spatial distributions in the brain of such hubs, for each group, while the corresponding anatomical information is reported in Supplementary Table~\ref{tab:S2}. In all cases we found hub regions peculiar for each group and hubs regions that are common to both groups. While significant differences are not observed between networks built from conventional approaches, hubs from multiplex analysis constitute a distinct set. 

In both conventional networks, healthy-specific hubs are located in medial superior frontal, lateral frontal cortices and thalamus, whereas schizophrenic-specific hubs are localized in posterior parts of the brain, such as cuneus, precuneus, and superior temporal cortices. Hubs shared by both groups are identified along the midline of the brain, in particular the medial superior frontal, precuneus and cingulate cortices. In the multiplex network, healthy-specific hubs controls are located in anterior cingulate, superior frontal, insula and superior temporal cortices, whereas those pertaining to schizophrenic patients are distributed over frontal, parietal and occipital cortices. Hub regions shared by both groups groups are localized in frontal, occipital cortices and cerebellum. Notably, no hub region has been identified in the precuneus cortex, a region well known to function as a hub in healthy individuals~\cite{van2013network}.

\begin{figure}[!t]
\centering
\includegraphics[width=9cm]{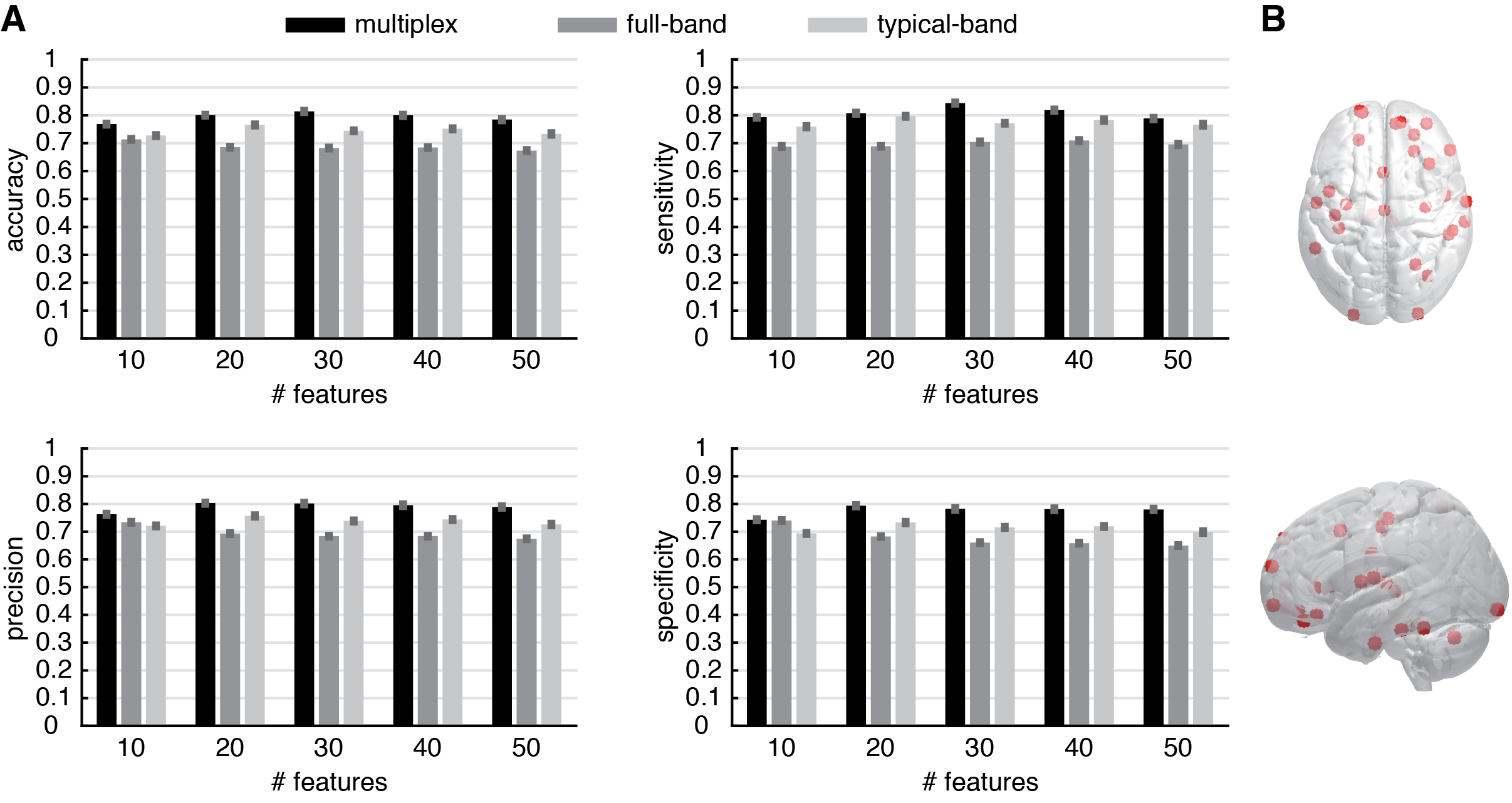}
\caption{\textbf{Discrimination performance of the multiplex functional networks vs. conventional networks.} (A) The statistical indicators of the discrimination between control and patient groups obtained from conventional network approaches (i.e. full-band and typical-band networks) are compared against the full multiplex functional network, which provides better overall discrimination. Note that the features are ROIs and their values are centrality scores. (B) Location of top 30 discriminating ROIs, obtained from multiplex analysis.}\label{fig:rforest}
\end{figure}

\section*{Discussion}

Resting state functional connectivity has been widely investigated with fMRI in the past two decades. Since the first study conducted by Biswal et al.~\cite{biswal1995functional}, functional connectivity has been defined as an inter-regional temporal correlation of fMRI signals that are preprocessed with band-pass filters, removing frequency components below 0.01 and above 0.1~Hz. In fact, the power spectrum of spontaneous fluctuations of fMRI signals roughly follows a $1/f$ power-law scaling~\cite{he2011scale}, where powers in the higher frequency range are relatively weaker than lower ones, suggesting the hypothesis that only the lower frequency range substantially contributes to brain's function. However, recent studies have reported that conventionally excluded frequency bands might provide additional insights on brain activity~\cite{bassett2006adaptive,liao2013functional,sasai2014frequency,thompson2015frequency,chen2015bold}. As a consequence, brain functional networks exhibit a peculiar architecture, consisting of a few regions acting as hubs, strongly dependent on the frequency components of brain activity that contribute to inter-regional interactions. However, a rigorous method to identify such hubs in networks built from different frequency bands simultaneously is a challenging problem remaining largely unexplored. 

Our results, based on multiplex modeling and analysis of the brain activity, provide convincing evidence that  characterization of brain functional networks can not prescind from considering the whole information observed from different frequency bands, simultaneously. This crucial finding allows to exploit new theoretical and computational tools for the analysis of brain activity and opens a new direction towards a deeper understanding of brain function and its operated integration. As a first hint of the power of the new methodology, we have shown that multiplex characterization of brain regions, in terms of network centrality, allows to find new areas of the brain that have never been classified as relevant in brain's functional integration (or the opposite). This is the case of ROIs in the precuneus cortex, a well-known region of highly central functional hubs ~\cite{van2013network}, that are not found by our multiplex network analysis, reflecting the importance of considering the whole information simultaneously, rather than aggregating or neglecting part of it~\cite{dedomenico2015ranking}.

We wondered if this result could be exploited for practical applications, where the choice of specific frequency bands might play a crucial role. We focused our attention on characterizing brain disorders in schizophrenic patients, a research topic of great interest that has been largely explored~\cite{bassett2008hierarchical,lynall2010functional,van2013abnormal}, although individual diagnosis based on brain imaging remains still undeveloped~\cite{rubinov2013schizophrenia}. 
With the aid of the MRI technique, it has been recently shown that regions affected by schizophrenia are distributed across the brain~\cite{glahn2008meta,ellison2009meta}, impelling researchers to move from the conventional perspective where the causes of disorders are localized in specific areas, to a wider perspective with emphasis on abnormality in brain structural and functional connectivity~\cite{van2014brain}. Studies on structural connectivity provided evidences that schizophrenic brains exhibit abnormal network architecture, characterized by reduced hierarchical organization, the loss of frontal hubs with emergence of non-frontal hubs~\cite{bassett2008hierarchical,lynall2010functional} and degraded rich-club organization~\cite{van2013abnormal}.  Methods not based on networks were able to provide satisfactory performance in discriminating schizophrenic patients from the analysis of their brain activity~\cite{yang2010hybrid,chyzhyk2015computer}, although they are often based on very complicated machine learning algorithms and make use of heterogenous data sources, thus not improving our understanding of brain function. Here, we have found that multiplex centrality profile of brain regions allow to discriminate between control and schizophrenic groups of individuals more accurately than centrality profiles calculated from networks obtained by using conventional approaches, such as aggregating and/or disregarding the measured activity. Nevertheless, the discrimination accuracy is comparable to other methods, with the additional advantage of providing a framework facilitating the interpretation of results, without relying on external data sources or phenotypic information. In fact, we were able to identify many regions distinctive of schizophrenic brains, some of them localized where abnormality has been previously suggested~\cite{rubinov2013schizophrenia,honea2005regional}. The analysis of dissimilarities between networks corresponding to different layers of the multiplex functional network, confirmed significant differences between healthy and schizophrenic individuals in specific frequency ranges, including the higher ones. This finding demonstrates that brain activity in higher frequencies provides unique information about functional interaction in the brain, even if their amplitudes are under-represented in the power spectrum.

The proposed methodology suggests a guideline for future studies designed to consider brain's inter-regional interactions at different frequencies, encouraging the application of other multiplex network measures to functional networks obtained, for instance, from variable brain states.

\section*{Acknowledgments}
M.D.D. acknowledges financial support from the Spanish program Juan de la Cierva (IJCI-2014-20225). SS was supported by JSPS Postdoctoral Fellowships for Research Abroad. A.A. acknowledges financial support from ICREA Academia and James S.\ McDonnell Foundation and Spanish MINECO FIS2015-71582.

\section*{Methods}

\heading{Overview of the data set and fMRI preprocessing} The publicly available MR data set contributed by The Center for Biomedical Research Excellence (COBRE) was used in this study. The data set was downloaded from the following repository: \url{http://fcon_1000.projects.nitrc.org/indi/retro/cobre.html}. It includes functional and anatomical MRI data acquired from 72 Schizophrenic patients and 75 healthy controls (age: 18--65 for both groups). One patient?s data was discarded from all analyses due to the shortness of the data length. The following pre-processing steps were applied to functional MR images by using the SPM8 package (Wellcome Department of Imaging Neuroscience, London, UK): motion-correction, slice-timing-correction, spatial smoothing with Gaussian kernel (5-mm full-width-at-half-maximum) and spatial normalization. Signal fluctuations of fMRI are driven by not only neural but also physiological effects -- such as respiration and cardiac pulsation -- and environmental conditions -- such as scanner instabilities and subject motion. These nuisance effects can be canceled out by discarding, for instance, the signal from the ROI centered in the white matter, the signal from the ventricular ROI, and the signal from the ROI located within the soft-tissue. We have linearly removed these components after temporally shifting them by optimal time-lags yielding the highest correlation with the averaged signal of all gray matter voxels~\cite{anderson2011network}.

\heading{Connectivity matrices} A set of 264 spherical ROIs (5~mm radii) was used to extract the mean signal within each ROI. For each individual, the coherence between all pairs of in-ROI averaged signals was estimated in specific frequency bands, as described in the text. We kept the edges between pairs of ROIs whose weight was significantly different from a null model where observed signals were replaced by surrogates. More specifically, we used the well-known iterative amplitude-adjusted Fourier transform (IAAFT) algorithm to build surrogate time series preserving the power spectrum and the probability density of the original ones, while removing higher-order self-correlations. For each pair of ROIs $i$ and $j$, we have verified that the distribution of the weights obtained from the null model corresponds to a Gaussian described by sample mean $\mu_{ij}$ and variance $\sigma^{2}_{ij}$. Let $w_{ij}$ indicate the weight obtained from empirical data: we have calculated the absolute Z score as $z_{ij}=|w_{ij}-\mu_{ij}|/\sigma_{ij}$ and discarded all those edges for which $z_{ij}<3$, corresponding to cross-coherence not statistically significant. We used the values $z_{ij}$ as entries of the resulting connectivity matrix.

\heading{Multiplex network model} A multilayer network allows to encode different types of interactions or relationships among a set of nodes. More specifically, in the case of our study we make use of multiplex networks to model functional connectivity. In a multiplex network, the links are of different type: one can assign a different ``color'' to each type, thus obtained an edge-colored representation of the network. In this type of architectures, nodes exist in one or more layers, i.e. it is not required that all nodes exist in all layers. Correlation networks, as the ones used in this study, define edge-colored graphs where each layer encodes the correlations observed in a specific frequency band. However, it has been shown that by interconnecting nodes with their replicas across layers, the resulting interconnected multiplex network can be described by an adjacency tensor~\cite{dedomenico2013mathematical} with components $M^{i\alpha}_{j\beta}$, an object generalizing the well-known concept of adjacency matrix to higher orders, encoding connections between node $i$ in layer $\alpha$ and node $j$ in layer $\beta$. For interconnected multiplex networks, $M^{i\alpha}_{j\beta}=0$ for $i\neq j$ and simultaneously $\alpha\neq\beta$. The presence of interconnections allows to exploit tensorial algebra to generalize many single-layer network descriptors, from centrality~\cite{dedomenico2015ranking} to mesoscale structure~\cite{mucha2010community,dedomenico2015identifying}. However, it is not always possible to assign a weight to inter-layer links by using the data, and it is common to parameterize the intensity of interconnections~\cite{gomez2013diffusion,dedomenico2014navigability}, i.e. $M^{i\alpha}_{j\beta}=D$ for $i=j$ and simultaneously $\alpha\neq\beta$, to study the resulting interconnected multiplex network as a function of this parameter $D$. This is exactly the case of the present study, where the choice of $D$ depends on the analysis of interest.

\heading{Structural reducibility of brain multiplex functional network} The analysis of structural reducibility of a multilayer network allows to find layers that provide redundant topological information, suggesting how to merge some layers with other ones, to obtain an optimal multilayer network~\cite{dedomenico2015structural}. The whole procedure can be summarized as follows: i) compute the distance (based on quantum Jensen-Shannon divergence) between all pairs of layers; ii) perform hierarchical
clustering of layers using such distance matrix and use changes in the relative entropy $q(\bullet)$ as the quality function for the resulting partition; iii) finally, choose the partition which maximizes the quality function, i.e. the distinguishability from the fully aggregated graph obtained by summing up the adjacency matrices of all layers. It is worth remarking that this analysis is independent on the choice of interconnections weight, i.e. it does not depend on $D$. Here, we do not enter into the details of the whole method; instead we focus on the Jensen-Shannon distance, that is a key measure for two analyses presented in this study. 

The components $A^{[\alpha]}_{ij}$ ($i,j=1,2,...,N$; being $N$ the number of ROIs in this study) of the adjacency matrix $\bm{A}^{[\alpha]}$ -- encoding layer $\alpha$ -- are obtained from the components of the multilayer adjacency tensor as $A^{[\alpha]}_{ij}=M^{i\alpha}_{j\alpha}$. Here, $A^{[\alpha]}_{ij}>0$ if there is correlation between ROIs $i$ and $j$ in the frequency band represented by $\alpha$. The Von Neumann entropy~\cite{braunstein2006laplacian,passerini2010quantifying} of the corresponding complex network is defined by
\begin{equation}
  h_{A^{[\alpha]}} = -\text{Tr}\left[\bm{\mathcal{L}}^{[\alpha]} \log_2 \bm{\mathcal{L}}^{[\alpha]}\right],
  \label{eq:h_laplacian}
\end{equation}
where $\bm{\mathcal{L}}^{[\alpha]}=c\times(\bm{S}^{[\alpha]}-\bm{A}^{[\alpha]})$ is the combinatorial Laplacian rescaled by $c = 1/\left(\sum\limits_{i,j=1}^{N} A_{ij}^{[\alpha]}\right)$, and $S$ is the diagonal matrix of the strengths of the nodes. From the eigen-decomposition of the Laplacian, it is possible to show that the entropy can be calculated by
\begin{equation}
  h_{A^{[\alpha]}} = -\sum\limits_{i=1}^{N} \lambda_i^{[\alpha]} \log_2 (\lambda_i^{[\alpha]}),
  \label{eq:h_eigenvalues}
\end{equation}
where $\{\lambda_1^{[\alpha]}, \lambda_2^{[\alpha]}, \ldots, \lambda_N^{[\alpha]}\}$ are the eigenvalues of $\bm{\mathcal{L}}^{[\alpha]}$.

The similarity of two layers can be calculated in terms of differences in their entropy. Given two rescaled Laplacian matrices $\bm{\mathcal{L}}^{[\alpha]}$ and $\bm{\mathcal{L}}^{[\beta]}$, it is possible to quantify to which extent layer $\alpha$ is different from layer $\beta$ by their Kullback-Liebler divergence, defined by
\begin{equation}
  \mathcal{D}_{KL}(\bm{\mathcal{L}}^{[\alpha]}||\bm{\mathcal{L}}^{[\beta]}) =
  \text{Tr}[\bm{\mathcal{L}}^{[\alpha]}(\log_2(\bm{\mathcal{L}}^{[\alpha]}) - \log_2(\bm{\mathcal{L}}^{[\beta]}))],
  \label{eq:KL}
\end{equation}
encoding the information gained about $\bm{\mathcal{L}}^{[\beta]}$ when the expectation is based only on $\bm{\mathcal{L}}^{[\alpha]}$. This divergence is not a metric and a more suitable dissimilarity measure is the Jensen-Shannon divergence, defined by
\begin{equation}
\label{eq:JSD}
\mathcal{D}_{\text{JS}}(\bm{\mathcal{L}}^{[\alpha]}||\bm{\mathcal{L}}^{[\beta]}) =
\frac{1}{2}\mathcal{D}_{\text{KL}}(\bm{\mathcal{L}}^{[\alpha]}||\bm{\mathcal{L}}^{[\mu]}) +
\frac{1}{2}\mathcal{D}_{\text{KL}}(\bm{\mathcal{L}}^{[\beta]}||\bm{\mathcal{L}}^{[\mu]}),
\end{equation}
where $\bm{\mathcal{L}}^{[\mu]}=\frac{1}{2}(\bm{\mathcal{L}}^{[\alpha]} + \bm{\mathcal{L}}^{[\beta]})$. It can be shown that $\sqrt{\mathcal{D}_{JS}}$, usually called Jensen--Shannon distance,
takes values in $[0,1]$, satisfies all the properties of a metric distance and provides a very powerful measure of dissimilarity between layers.

\heading{Random forest classification} Machine learning has been used to train a classifier to distinguish between control and schizophrenic individuals. We used the random forest classifier~\cite{breiman2001random}, well-known for its robustness and for facilitating the interpretation of results. We have fixed to 5 the maximum number of terminal nodes trees the forest can have and to 2 the number of variables randomly sampled as candidates at each split. We have verified that forests consisting of 700 trees where enough to reach stable results within this setup.

Given the importance of interconnections weight for our analysis and, at the same time, the lack of knowledge about its value, we used random forest to learn also which value of $D$ would be more suitable for calculations. 

We have performed a first exploratory classification using a leave-one-out approach to maximize the amount of data used for training the classifier. The result of each classification, corresponding to exactly one different individual (without replacement) left out, was accompanied by the importance assigned by the classifier to each ROI in terms of mean decrease in its Gini index. Therefore, for each individual and each value of $D$, we have ranked the ROIs according to this measure and, eventually, summed up the ranks corresponding to all classifications. 

The result of the exploratory classification was an overall ranking suggesting which ROIs, in general, have been more crucial than others in the classification process. Therefore, we performed a second classification round by using only the top ROIs according to the above ranking. We first varied the number of kept features and the value of $D$, to find the values with best classification performances in terms of accuracy (see Supplementary Fig.~\ref{fig:SI1}). The numerical analysis indicated that the best classification is achieved for interconnections weight close to $24.7708$ and about 30 top ROIs: that value of $D$ and that sub-set of ROIs have been used for analysis reported in the text.

Using a similar approach, we have compared the best performance obtained from the full multiplex functional network (12 layers) against multiplex functional networks obtained by keeping layers in the typical band (Supplementary Fig.~\ref{fig:SI2}) and against  classifier trained by including phenotypic data (Supplementary Fig.~\ref{fig:SI3}). In all cases, the classification obtained using the full multiplex functional network was equal or better than the other ones. 


\heading{ROIs PageRank centrality} PageRank is a measure of node's centrality originally introduced by Google founders to rank Web pages according to their importance in the Internet~\cite{brin1998,erman2015google}. The algorithm consider a random walker exploring the network with the following rules: 85\% of times the walker jumps from the current node to one chosen with uniform probability from the neighborhood, whereas 15\% of times the walker is allowed to jump to any node of the network, with uniform probability. The stationary probability of finding the walker in a specific node is then used to rank the importance of nodes in the network, the rationale being that central nodes have high number of incoming links from other important nodes.

The natural extension of the PageRank algorithm to the context of multiplex networks has been recently introduced~\cite{dedomenico2015ranking} and proved to perform better than its single-layer counterpart in some applications. Let us indicate with $R^{i\alpha}_{j\beta}$ the transition tensor, governing the dynamics of a random walker jumping to neighboring nodes with rate $0.85$ and teleporting to any other node in the network with rate $0.15$. This rank-4 tensor is given by
\begin{eqnarray}\label{RWPageRank}
	R^{i\alpha}_{j\beta} = 0.85\times T^{i\alpha}_{j\beta} + \frac{0.15}{NL}u^{i\alpha}_{j\beta},
\end{eqnarray}
where $T^{i\alpha}_{j\beta}$ governs the standard moves of a classical random walker from a node $i$ in layer $\alpha$ to one of its neighbors $j$ in layer $\beta$, $L$ is the total number of layers and $u^{i\alpha}_{j\beta}$ is the rank-4 tensor with all components equal to 1. The steady-state solution of the master equation 
\begin{eqnarray}
\pi_{j\beta}(t+1) =\sum_{i=1}^{N}\sum_{\alpha=1}^{L} R^{i\alpha}_{j\beta}\pi_{i\alpha}(t),
\end{eqnarray}
obtained in the limit $t\longrightarrow\infty$, provides the PageRank centrality for interconnected multiplex networks. To compute the overall PageRank of a node, accounting for the whole interconnected topology, we can safely sum up the stationary probabilities $\pi_{j\beta}^{\star}$ over the layers, to obtain the components of the centrality profile vector $\tilde{\pi}^{\star}_{j}=\sum\limits_{\beta=1}^{L}\pi_{j\beta}^{\star}$ used in our analysis. It is worth remarking that the interconnection weight used for this purpose is $D=24.7708$, the one yielding the highest classification accuracy.


\renewcommand*{\thefigure}{{\bf \arabic{figure}}}
\renewcommand{\figurename}{{\bf Supplementary Figure}}
\renewcommand*{\thetable}{{\bf \arabic{table}}}
\renewcommand{\tablename}{{\bf Supplementary Table}}

\makeatletter
\setcounter{figure}{0}

\begin{figure}[!h]
\centering
\includegraphics[width=9cm]{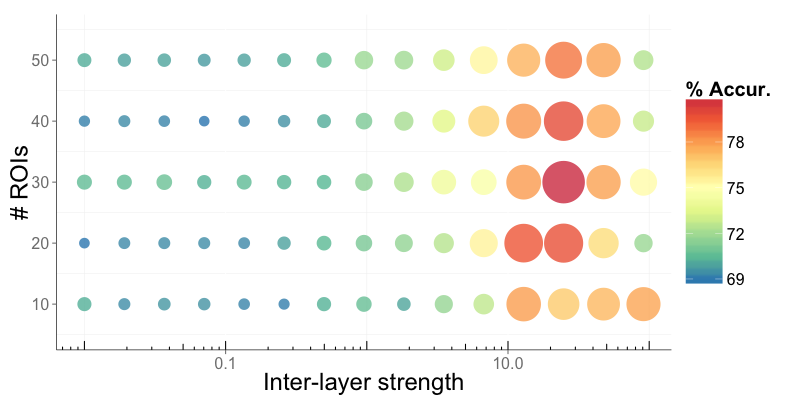}
\caption{\textbf{Maximizing classification accuracy.} Layers in the multiplex functional network are interconnected to allow the calculation of layer's and node's properties. However, the weight of such inter-layer links can not be deduced from the data. At the same time, not all ROIs have enough discriminant power and using sub-sets of them reduces the noise and generally improves the classification. We have chosen the value of the inter-layer strength ($D = 24.7708$) and the size of an appropriate sub-set of ROIs (30) where maximum classification accuracy is achieved (see Methods for details).}\label{fig:SI1}
\end{figure}

\begin{figure}[!h]
\centering
\includegraphics[width=9cm]{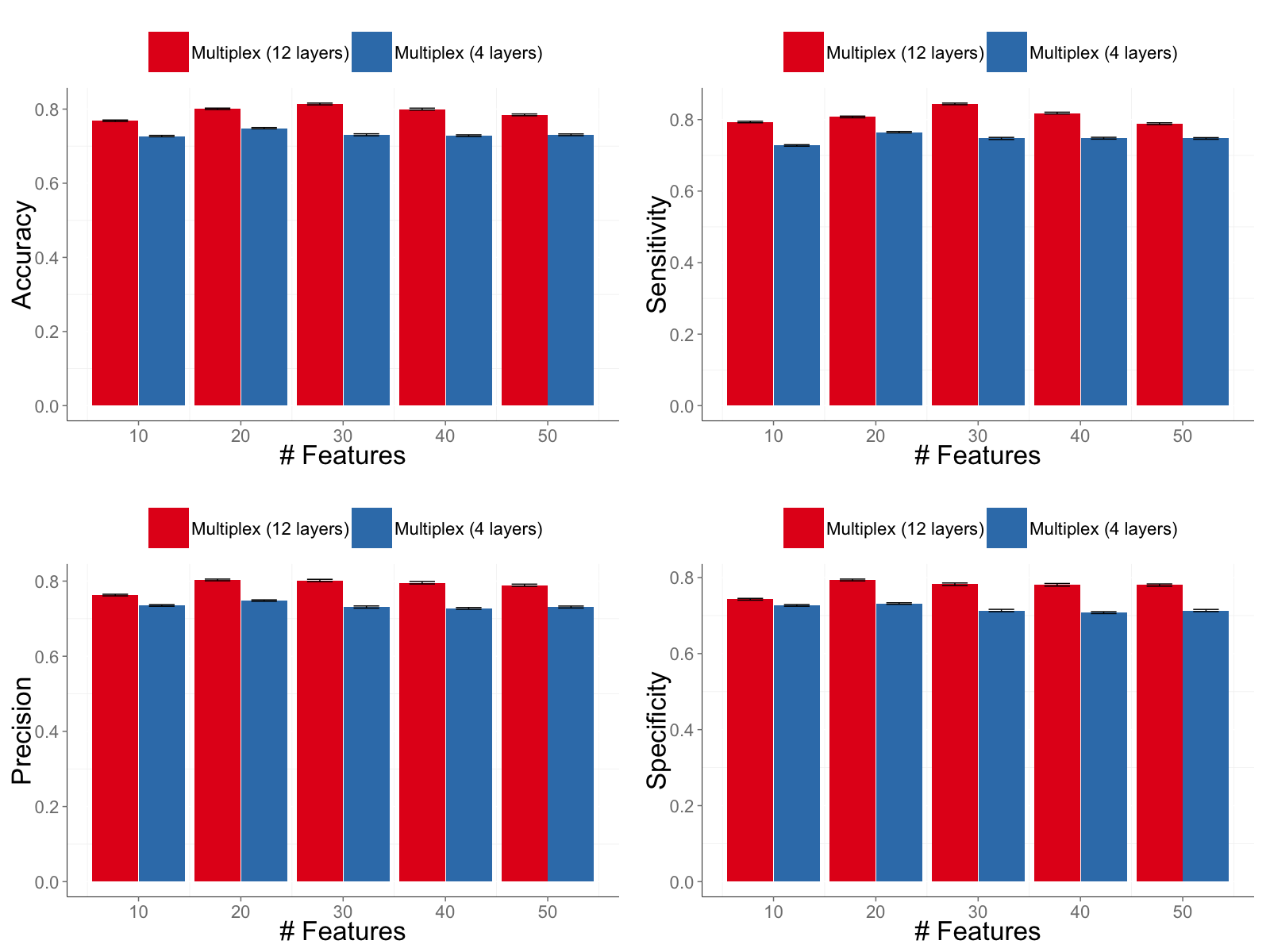}
\caption{\textbf{Discrimination performance of two different multiplex functional networks.} A new multiplex network, consisting only of the layers corresponding to the typical frequency range, has been built and used for discriminating between control and patient groups. The statistical indicators of the discrimination are compared against the full multiplex functional network discussed in the text, providing better overall discrimination. Bars indicate standard errors.}\label{fig:SI2}
\end{figure}

\begin{figure}[!h]
\centering
\includegraphics[width=9cm]{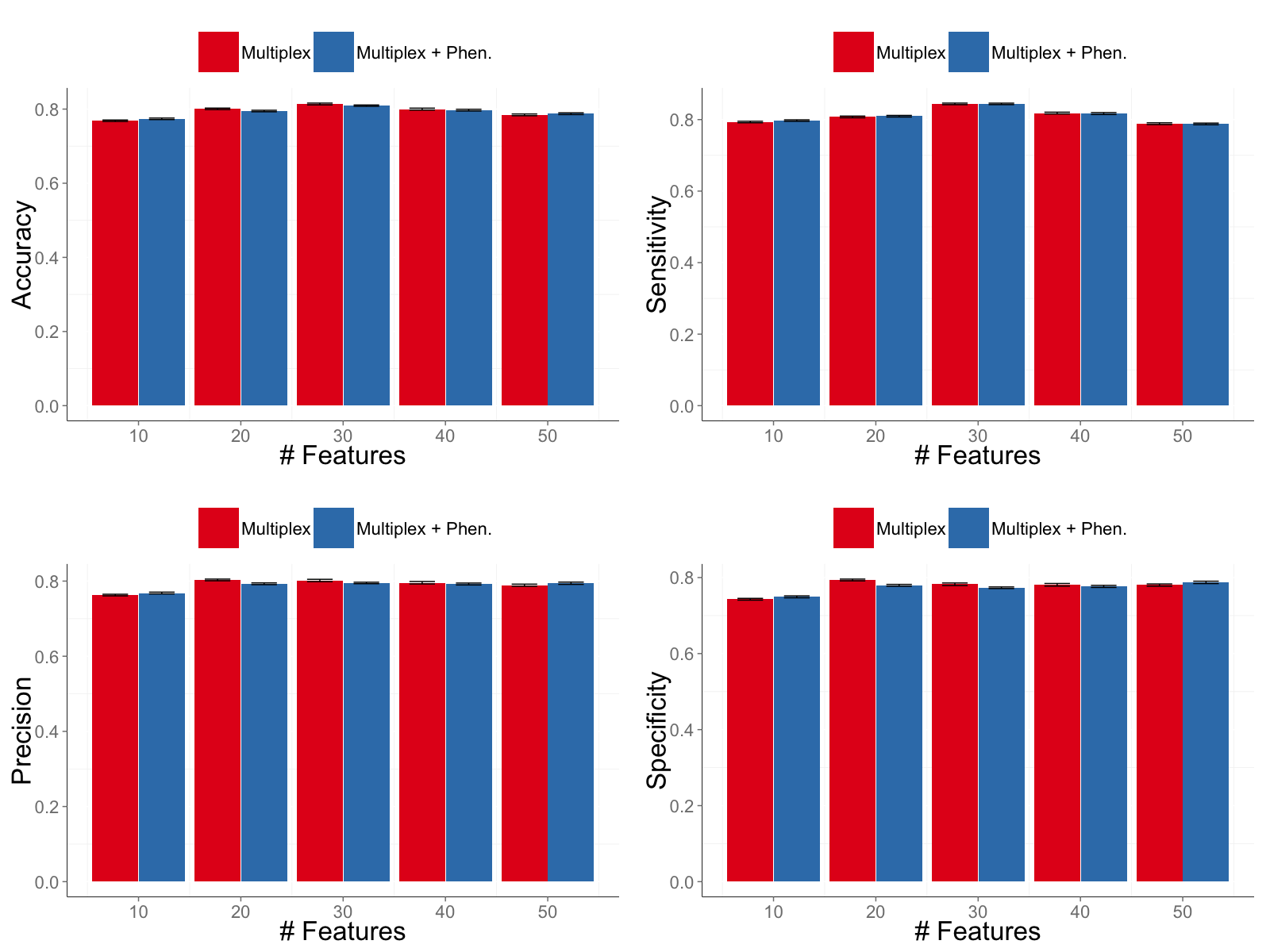}
\caption{\textbf{Discrimination performance of the multiplex functional networks with and without phenotypic information.} The statistical indicators of the discrimination between control and patient groups obtained from the full multiplex functional network before and after including phenotypic information in the machine learning process. The results are not significantly different, indicating that phenotypic data is redundant in this case and does not improve discrimination. Bars indicate standard errors.
}\label{fig:SI3}
\end{figure}

\clearpage
\newpage

\begin{table}[!h]
\centering
\caption{}
\label{tab:S1}
\begin{footnotesize}
\renewcommand{\arraystretch}{0.3}
\begin{tabular}{@{}lllllllll@{}}
\toprule
\multicolumn{1}{c}{\textbf{Network model}} & \multicolumn{1}{c}{\textbf{Group}} & \multicolumn{3}{c}{\textbf{\begin{tabular}[c]{@{}c@{}}MNI \\ coordinate\end{tabular}}} & \multicolumn{3}{c}{\textbf{\begin{tabular}[c]{@{}c@{}}Talairach \\ coordinate\end{tabular}}} & \multicolumn{1}{c}{\textbf{Anatomical label}} \\ \midrule
                                           &                                    & \textbf{X}                  & \textbf{Y}                  & \textbf{Z}                 & \textbf{X}                    & \textbf{Y}                    & \textbf{Z}                   &                                               \\
multiplex                                  & healthy                            & 58                          & -16                         & 7                          & 53                            & -17                           & 9                            & Transverse Temporal Gyrus                     \\
                                           &                                    & 32                          & -26                         & 13                         & 29                            & -27                           & 14                           & Claustrum                                     \\
                                           &                                    & -3                          & 42                          & 16                         & -4                            & 37                            & 21                           & Anterior Cingulate                            \\
                                           &                                    & 0                           & 30                          & 27                         & -1                            & 25                            & 30                           & Cingulate Gyrus                               \\
                                           &                                    & 34                          & 16                          & -8                         & 31                            & 14                            & -2                           & Claustrum                                     \\
                                           &                                    & -28                         & 52                          & 21                         & -27                           & 46                            & 26                           & Superior Frontal Gyrus                        \\
                                           &                                    & 32                          & 14                          & 56                         & 29                            & 7                             & 55                           & Middle Frontal Gyrus                          \\
                                           &                                    & -11                         & 45                          & 8                          & -11                           & 40                            & 14                           & Medial Frontal Gyrus                          \\
                                           &                                    & -13                         & -17                         & 75                         & -14                           & -23                           & 69                           & Precentral Gyrus                              \\
                                           &                                    &                             &                             &                            &                               &                               &                              &                                               \\
                                           & schizophrenia                      & -16                         & -5                          & 71                         & -17                           & -11                           & 67                           & Superior Frontal Gyrus                        \\
                                           &                                    & 24                          & -87                         & 24                         & 21                            & -85                           & 19                           & Cuneus                                        \\
                                           &                                    & 19                          & -8                          & 64                         & 16                            & -14                           & 61                           & Superior Frontal Gyrus                        \\
                                           &                                    & -3                          & 26                          & 44                         & -4                            & 20                            & 45                           & Medial Frontal Gyrus                          \\
                                           &                                    & 46                          & -59                         & 4                          & 42                            & -57                           & 3                            & Middle Temporal Gyrus                         \\
                                           &                                    & -7                          & -21                         & 65                         & -8                            & -26                           & 60                           & Medial Frontal Gyrus                          \\
                                           &                                    & 34                          & 38                          & -12                        & 31                            & 35                            & -4                           & Inferior Frontal Gyrus                        \\
                                           &                                    & -47                         & -76                         & -10                        & -45                           & -72                           & -12                          & Fusiform Gyrus                                \\
                                           &                                    & -37                         & -29                         & -26                        & -36                           & -27                           & -22                          & Parahippocampal Gyrus                         \\
                                           &                                    &                             &                             &                            &                               &                               &                              &                                               \\
                                           & common                             & 11                          & -54                         & 17                         & 9                             & -53                           & 15                           & Posterior Cingulate                           \\
                                           &                                    & 17                          & -80                         & -34                        & 15                            & -74                           & -33                          & Uvula                                         \\
                                           &                                    & -50                         & -7                          & -39                        & -48                           & -5                            & -32                          & Inferior Temporal Gyrus                       \\
                                           &                                    & -32                         & -1                          & 54                         & -31                           & -6                            & 52                           & Precentral Gyrus                              \\
                                           &                                    &                             &                             &                            &                               &                               &                              &                                               \\
full-band                                  & healthy                            & -60                         & -25                         & 14                         & -57                           & -26                           & 14                           & Superior Temporal Gyrus                       \\
                                           &                                    & -7                          & -52                         & 61                         & -8                            & -55                           & 54                           & Precuneus                                     \\
                                           &                                    & 58                          & -16                         & 7                          & 53                            & -17                           & 9                            & Transverse Temporal Gyrus                     \\
                                           &                                    & 6                           & -72                         & 24                         & 4                             & -71                           & 20                           & Cuneus                                        \\
                                           &                                    & 0                           & 30                          & 27                         & -1                            & 25                            & 30                           & Cingulate Gyrus                               \\
                                           &                                    & -16                         & -77                         & 34                         & -16                           & -76                           & 28                           & Cuneus                                        \\
                                           &                                    & -3                          & -81                         & 21                         & -4                            & -79                           & 16                           & Cuneus                                        \\
                                           &                                    & 10                          & -62                         & 61                         & 8                             & -64                           & 54                           & Superior Parietal Lobule                      \\
                                           &                                    &                             &                             &                            &                               &                               &                              &                                               \\
                                           & schizophrenia                      & -10                         & -18                         & 7                          & -11                           & -19                           & 8                            & Thalamus                                      \\
                                           &                                    & -20                         & 64                          & 19                         & -20                           & 57                            & 25                           & Superior Frontal Gyrus                        \\
                                           &                                    & 26                          & 50                          & 27                         & 23                            & 43                            & 32                           & Superior Frontal Gyrus                        \\
                                           &                                    & 27                          & -97                         & -13                        & 24                            & -92                           & -15                          & Fusiform Gyrus                                \\
                                           &                                    & -55                         & -9                          & 12                         & -53                           & -10                           & 13                           & Precentral Gyrus                              \\
                                           &                                    & 36                          & 22                          & 3                          & 33                            & 19                            & 8                            & Insula                                        \\
                                           &                                    & 9                           & -4                          & 6                          & 7                             & -6                            & 9                            & Thalamus                                      \\
                                           &                                    & 47                          & -30                         & 49                         & 42                            & -33                           & 46                           & Inferior Parietal Lobule                      \\
                                           &                                    &                             &                             &                            &                               &                               &                              &                                               \\
                                           & common                             & 0                           & -15                         & 47                         & -1                            & -19                           & 45                           & Paracentral Lobule                            \\
                                           &                                    & -3                          & 2                           & 53                         & -4                            & -3                            & 51                           & Medial Frontal Gyrus                          \\
                                           &                                    & 4                           & -48                         & 51                         & 2                             & -50                           & 46                           & Precuneus                                     \\
                                           &                                    & -3                          & 26                          & 44                         & -4                            & 20                            & 45                           & Medial Frontal Gyrus                          \\
                                           &                                    & -1                          & 15                          & 44                         & -2                            & 10                            & 44                           & Medial Frontal Gyrus                          \\
                                           &                                    &                             &                             &                            &                               &                               &                              &                                               \\
typical-band                               & healthy                            & 58                          & -16                         & 7                          & 53                            & -17                           & 9                            & Transverse Temporal Gyrus                     \\
                                           &                                    & -60                         & -25                         & 14                         & -57                           & -26                           & 14                           & Superior Temporal Gyrus                       \\
                                           &                                    & 6                           & -72                         & 24                         & 4                             & -71                           & 20                           & Cuneus                                        \\
                                           &                                    & 11                          & -39                         & 50                         & 9                             & -42                           & 46                           & Precuneus                                     \\
                                           &                                    & 10                          & -62                         & 61                         & 8                             & -64                           & 54                           & Superior Parietal Lobule                      \\
                                           &                                    & 8                           & -72                         & 11                         & 6                             & -70                           & 8                            & Cuneus                                        \\
                                           &                                    & 32                          & -26                         & 13                         & 29                            & -27                           & 14                           & Claustrum                                     \\
                                           &                                    &                             &                             &                            &                               &                               &                              &                                               \\
                                           & schizophrenia                      & -10                         & -18                         & 7                          & -11                           & -19                           & 8                            & Thalamus                                      \\
                                           &                                    & -5                          & 18                          & 34                         & -6                            & 13                            & 35                           & Cingulate Gyrus                               \\
                                           &                                    & 26                          & 50                          & 27                         & 23                            & 43                            & 32                           & Superior Frontal Gyrus                        \\
                                           &                                    & 7                           & 8                           & 51                         & 5                             & 2                             & 50                           & Medial Frontal Gyrus                          \\
                                           &                                    & 5                           & 23                          & 37                         & 3                             & 17                            & 39                           & Cingulate Gyrus                               \\
                                           &                                    & 0                           & 30                          & 27                         & -1                            & 25                            & 30                           & Cingulate Gyrus                               \\
                                           &                                    & -55                         & -9                          & 12                         & -53                           & -10                           & 13                           & Precentral Gyrus                              \\
                                           &                                    &                             &                             &                            &                               &                               &                              &                                               \\
                                           & common                             & -7                          & -52                         & 61                         & -8                            & -55                           & 54                           & Precuneus                                     \\
                                           &                                    & 0                           & -15                         & 47                         & -1                            & -19                           & 45                           & Paracentral Lobule                            \\
                                           &                                    & -3                          & 2                           & 53                         & -4                            & -3                            & 51                           & Medial Frontal Gyrus                          \\
                                           &                                    & 4                           & -48                         & 51                         & 2                             & -50                           & 46                           & Precuneus                                     \\
                                           &                                    & -3                          & 26                          & 44                         & -4                            & 20                            & 45                           & Medial Frontal Gyrus                          \\
                                           &                                    & -1                          & 15                          & 44                         & -2                            & 10                            & 44                           & Medial Frontal Gyrus                          \\ \bottomrule
\end{tabular}
\end{footnotesize}
\end{table}

\clearpage
\newpage
\begin{table}[]
\centering
\caption{}
\label{tab:S2}
\begin{footnotesize}
\renewcommand{\arraystretch}{0.6}
\begin{tabular}{@{}lllllll@{}}
\toprule
\multicolumn{3}{c}{\textbf{\begin{tabular}[c]{@{}c@{}}MNI\\ coordinate\end{tabular}}} & \multicolumn{3}{c}{\textbf{\begin{tabular}[c]{@{}c@{}}Talairach\\ coordinate\end{tabular}}} & \multicolumn{1}{c}{\textbf{Anatomical label}} \\ \midrule
X                           & Y                          & Z                          & X                             & Y                            & Z                            &                                               \\
27                          & -97                        & -13                        & 24                            & -92                          & -15                          & Fusiform Gyrus                                \\
-10                         & -18                        & 7                          & -11                           & -19                          & 8                            & Thalamus                                      \\
52                          & -34                        & -27                        & 48                            & -32                          & -23                          & Fusiform Gyrus                                \\
36                          & 22                         & 3                          & 33                            & 19                           & 8                            & Insula                                        \\
-25                         & -98                        & -12                        & -25                           & -92                          & -15                          & Fusiform Gyrus                                \\
65                          & -24                        & -19                        & 60                            & -23                          & -15                          & Middle Temporal Gyrus                         \\
-40                         & -19                        & 54                         & -39                           & -23                          & 50                           & Postcentral Gyrus                             \\
6                           & -24                        & 0                          & 5                             & -24                          & 2                            & Thalamus                                      \\
33                          & -12                        & -34                        & 30                            & -10                          & -27                          & Uncus                                         \\
24                          & 32                         & -18                        & 22                            & 30                           & -10                          & Sub-Gyral                                     \\
34                          & 54                         & -13                        & 31                            & 50                           & -4                           & Middle Frontal Gyrus                          \\
55                          & -31                        & -17                        & 50                            & -30                          & -13                          & Inferior Temporal Gyrus                       \\
0                           & -15                        & 47                         & -1                            & -19                          & 45                           & Paracentral Lobule                            \\
66                          & -8                         & 25                         & 61                            & -11                          & 26                           & Precentral Gyrus                              \\
-45                         & 0                          & 9                          & -43                           & -2                           & 11                           & Insula                                        \\
-56                         & -45                        & -24                        & -53                           & -42                          & -22                          & Fusiform Gyrus                                \\
-55                         & -9                         & 12                         & -53                           & -10                          & 13                           & Precentral Gyrus                              \\
25                          & -58                        & 60                         & 22                            & -60                          & 53                           & Superior Parietal Lobule                      \\
-21                         & 41                         & -20                        & -20                           & 39                           & -11                          & Middle Frontal Gyrus                          \\
-31                         & -10                        & -36                        & -30                           & -8                           & -30                          & Uncus                                         \\
9                           & 54                         & 3                          & 8                             & 49                           & 10                           & Medial Frontal Gyrus                          \\
13                          & 55                         & 38                         & 11                            & 47                           & 42                           & Superior Frontal Gyrus                        \\
-20                         & 64                         & 19                         & -20                           & 57                           & 25                           & Superior Frontal Gyrus                        \\
35                          & -67                        & -34                        & 32                            & -62                          & -32                          & Cerebellar Tonsil                             \\
49                          & -3                         & -38                        & 45                            & -2                           & -30                          & Middle Temporal Gyrus                         \\
-37                         & -29                        & -26                        & -36                           & -27                          & -22                          & Parahippocampal Gyrus                         \\
24                          & 45                         & -15                        & 22                            & 42                           & -6                           & Medial Frontal Gyrus                          \\
-1                          & 15                         & 44                         & -2                            & 10                           & 44                           & Medial Frontal Gyrus                          \\
-18                         & 63                         & -9                         & -18                           & 59                           & 0                            & Medial Frontal Gyrus                          \\
53                          & 33                         & 1                          & 49                            & 29                           & 8                            & Inferior Frontal Gyrus                        \\ \bottomrule
\end{tabular}
\end{footnotesize}
\end{table}

\clearpage
\bibliographystyle{apsrev}
\bibliography{multiplex_brain}

\begin{thebibliography}{50}
\expandafter\ifx\csname natexlab\endcsname\relax\def\natexlab#1{#1}\fi
\expandafter\ifx\csname bibnamefont\endcsname\relax
  \def\bibnamefont#1{#1}\fi
\expandafter\ifx\csname bibfnamefont\endcsname\relax
  \def\bibfnamefont#1{#1}\fi
\expandafter\ifx\csname citenamefont\endcsname\relax
  \def\citenamefont#1{#1}\fi
\expandafter\ifx\csname url\endcsname\relax
  \def\url#1{\texttt{#1}}\fi
\expandafter\ifx\csname urlprefix\endcsname\relax\def\urlprefix{URL }\fi
\providecommand{\bibinfo}[2]{#2}
\providecommand{\eprint}[2][]{\url{#2}}

\bibitem[{\citenamefont{Bassett and Bullmore}(2006)}]{bassett2006small}
\bibinfo{author}{\bibfnamefont{D.~S.} \bibnamefont{Bassett}} \bibnamefont{and}
  \bibinfo{author}{\bibfnamefont{E.}~\bibnamefont{Bullmore}},
  \bibinfo{journal}{The neuroscientist} \textbf{\bibinfo{volume}{12}},
  \bibinfo{pages}{512} (\bibinfo{year}{2006}).

\bibitem[{\citenamefont{Bullmore and Sporns}(2009)}]{bullmore2009complex}
\bibinfo{author}{\bibfnamefont{E.}~\bibnamefont{Bullmore}} \bibnamefont{and}
  \bibinfo{author}{\bibfnamefont{O.}~\bibnamefont{Sporns}},
  \bibinfo{journal}{Nature Reviews Neuroscience} \textbf{\bibinfo{volume}{10}},
  \bibinfo{pages}{186} (\bibinfo{year}{2009}).

\bibitem[{\citenamefont{Bullmore and Sporns}(2012)}]{bullmore2012economy}
\bibinfo{author}{\bibfnamefont{E.}~\bibnamefont{Bullmore}} \bibnamefont{and}
  \bibinfo{author}{\bibfnamefont{O.}~\bibnamefont{Sporns}},
  \bibinfo{journal}{Nature Reviews Neuroscience} \textbf{\bibinfo{volume}{13}},
  \bibinfo{pages}{336} (\bibinfo{year}{2012}).

\bibitem[{\citenamefont{Van Den~Heuvel and Pol}(2010)}]{van2010exploring}
\bibinfo{author}{\bibfnamefont{M.~P.} \bibnamefont{Van Den~Heuvel}}
  \bibnamefont{and} \bibinfo{author}{\bibfnamefont{H.~E.~H.}
  \bibnamefont{Pol}}, \bibinfo{journal}{European Neuropsychopharmacology}
  \textbf{\bibinfo{volume}{20}}, \bibinfo{pages}{519} (\bibinfo{year}{2010}).

\bibitem[{\citenamefont{Poldrack and Farah}(2015)}]{poldrack2015progress}
\bibinfo{author}{\bibfnamefont{R.~A.} \bibnamefont{Poldrack}} \bibnamefont{and}
  \bibinfo{author}{\bibfnamefont{M.~J.} \bibnamefont{Farah}},
  \bibinfo{journal}{Nature} \textbf{\bibinfo{volume}{526}},
  \bibinfo{pages}{371} (\bibinfo{year}{2015}).

\bibitem[{\citenamefont{Achard et~al.}(2006)\citenamefont{Achard, Salvador,
  Whitcher, Suckling, and Bullmore}}]{achard2006resilient}
\bibinfo{author}{\bibfnamefont{S.}~\bibnamefont{Achard}},
  \bibinfo{author}{\bibfnamefont{R.}~\bibnamefont{Salvador}},
  \bibinfo{author}{\bibfnamefont{B.}~\bibnamefont{Whitcher}},
  \bibinfo{author}{\bibfnamefont{J.}~\bibnamefont{Suckling}}, \bibnamefont{and}
  \bibinfo{author}{\bibfnamefont{E.}~\bibnamefont{Bullmore}},
  \bibinfo{journal}{The Journal of Neuroscience} \textbf{\bibinfo{volume}{26}},
  \bibinfo{pages}{63} (\bibinfo{year}{2006}).

\bibitem[{\citenamefont{Power et~al.}(2013)\citenamefont{Power, Schlaggar,
  Lessov-Schlaggar, and Petersen}}]{power2013evidence}
\bibinfo{author}{\bibfnamefont{J.~D.} \bibnamefont{Power}},
  \bibinfo{author}{\bibfnamefont{B.~L.} \bibnamefont{Schlaggar}},
  \bibinfo{author}{\bibfnamefont{C.~N.} \bibnamefont{Lessov-Schlaggar}},
  \bibnamefont{and} \bibinfo{author}{\bibfnamefont{S.~E.}
  \bibnamefont{Petersen}}, \bibinfo{journal}{Neuron}
  \textbf{\bibinfo{volume}{79}}, \bibinfo{pages}{798} (\bibinfo{year}{2013}).

\bibitem[{\citenamefont{Cordes et~al.}(2001)\citenamefont{Cordes, Haughton,
  Arfanakis, Carew, Turski, Moritz, Quigley, and
  Meyerand}}]{cordes2001frequencies}
\bibinfo{author}{\bibfnamefont{D.}~\bibnamefont{Cordes}},
  \bibinfo{author}{\bibfnamefont{V.~M.} \bibnamefont{Haughton}},
  \bibinfo{author}{\bibfnamefont{K.}~\bibnamefont{Arfanakis}},
  \bibinfo{author}{\bibfnamefont{J.~D.} \bibnamefont{Carew}},
  \bibinfo{author}{\bibfnamefont{P.~A.} \bibnamefont{Turski}},
  \bibinfo{author}{\bibfnamefont{C.~H.} \bibnamefont{Moritz}},
  \bibinfo{author}{\bibfnamefont{M.~A.} \bibnamefont{Quigley}},
  \bibnamefont{and} \bibinfo{author}{\bibfnamefont{M.~E.}
  \bibnamefont{Meyerand}}, \bibinfo{journal}{American Journal of
  Neuroradiology} \textbf{\bibinfo{volume}{22}}, \bibinfo{pages}{1326}
  (\bibinfo{year}{2001}).

\bibitem[{\citenamefont{Cordes et~al.}(2002)\citenamefont{Cordes, Haughton,
  Carew, Arfanakis, and Maravilla}}]{cordes2002hierarchical}
\bibinfo{author}{\bibfnamefont{D.}~\bibnamefont{Cordes}},
  \bibinfo{author}{\bibfnamefont{V.}~\bibnamefont{Haughton}},
  \bibinfo{author}{\bibfnamefont{J.~D.} \bibnamefont{Carew}},
  \bibinfo{author}{\bibfnamefont{K.}~\bibnamefont{Arfanakis}},
  \bibnamefont{and}
  \bibinfo{author}{\bibfnamefont{K.}~\bibnamefont{Maravilla}},
  \bibinfo{journal}{Magnetic resonance imaging} \textbf{\bibinfo{volume}{20}},
  \bibinfo{pages}{305} (\bibinfo{year}{2002}).

\bibitem[{\citenamefont{Fox and Raichle}(2007)}]{fox2007spontaneous}
\bibinfo{author}{\bibfnamefont{M.~D.} \bibnamefont{Fox}} \bibnamefont{and}
  \bibinfo{author}{\bibfnamefont{M.~E.} \bibnamefont{Raichle}},
  \bibinfo{journal}{Nature Reviews Neuroscience} \textbf{\bibinfo{volume}{8}},
  \bibinfo{pages}{700} (\bibinfo{year}{2007}).

\bibitem[{\citenamefont{Sporns et~al.}(2007)\citenamefont{Sporns, Honey, and
  K{\"o}tter}}]{sporns2007identification}
\bibinfo{author}{\bibfnamefont{O.}~\bibnamefont{Sporns}},
  \bibinfo{author}{\bibfnamefont{C.~J.} \bibnamefont{Honey}}, \bibnamefont{and}
  \bibinfo{author}{\bibfnamefont{R.}~\bibnamefont{K{\"o}tter}},
  \bibinfo{journal}{PloS one} \textbf{\bibinfo{volume}{2}},
  \bibinfo{pages}{e1049} (\bibinfo{year}{2007}).

\bibitem[{\citenamefont{Bassett et~al.}(2008)\citenamefont{Bassett, Bullmore,
  Verchinski, Mattay, Weinberger, and
  Meyer-Lindenberg}}]{bassett2008hierarchical}
\bibinfo{author}{\bibfnamefont{D.~S.} \bibnamefont{Bassett}},
  \bibinfo{author}{\bibfnamefont{E.}~\bibnamefont{Bullmore}},
  \bibinfo{author}{\bibfnamefont{B.~A.} \bibnamefont{Verchinski}},
  \bibinfo{author}{\bibfnamefont{V.~S.} \bibnamefont{Mattay}},
  \bibinfo{author}{\bibfnamefont{D.~R.} \bibnamefont{Weinberger}},
  \bibnamefont{and}
  \bibinfo{author}{\bibfnamefont{A.}~\bibnamefont{Meyer-Lindenberg}},
  \bibinfo{journal}{The Journal of Neuroscience} \textbf{\bibinfo{volume}{28}},
  \bibinfo{pages}{9239} (\bibinfo{year}{2008}).

\bibitem[{\citenamefont{Lynall et~al.}(2010)\citenamefont{Lynall, Bassett,
  Kerwin, McKenna, Kitzbichler, Muller, and Bullmore}}]{lynall2010functional}
\bibinfo{author}{\bibfnamefont{M.-E.} \bibnamefont{Lynall}},
  \bibinfo{author}{\bibfnamefont{D.~S.} \bibnamefont{Bassett}},
  \bibinfo{author}{\bibfnamefont{R.}~\bibnamefont{Kerwin}},
  \bibinfo{author}{\bibfnamefont{P.~J.} \bibnamefont{McKenna}},
  \bibinfo{author}{\bibfnamefont{M.}~\bibnamefont{Kitzbichler}},
  \bibinfo{author}{\bibfnamefont{U.}~\bibnamefont{Muller}}, \bibnamefont{and}
  \bibinfo{author}{\bibfnamefont{E.}~\bibnamefont{Bullmore}},
  \bibinfo{journal}{The Journal of Neuroscience} \textbf{\bibinfo{volume}{30}},
  \bibinfo{pages}{9477} (\bibinfo{year}{2010}).

\bibitem[{\citenamefont{Rubinov and Sporns}(2010)}]{rubinov2010complex}
\bibinfo{author}{\bibfnamefont{M.}~\bibnamefont{Rubinov}} \bibnamefont{and}
  \bibinfo{author}{\bibfnamefont{O.}~\bibnamefont{Sporns}},
  \bibinfo{journal}{Neuroimage} \textbf{\bibinfo{volume}{52}},
  \bibinfo{pages}{1059} (\bibinfo{year}{2010}).

\bibitem[{\citenamefont{Zuo et~al.}(2012)\citenamefont{Zuo, Ehmke, Mennes,
  Imperati, Castellanos, Sporns, and Milham}}]{zuo2012network}
\bibinfo{author}{\bibfnamefont{X.-N.} \bibnamefont{Zuo}},
  \bibinfo{author}{\bibfnamefont{R.}~\bibnamefont{Ehmke}},
  \bibinfo{author}{\bibfnamefont{M.}~\bibnamefont{Mennes}},
  \bibinfo{author}{\bibfnamefont{D.}~\bibnamefont{Imperati}},
  \bibinfo{author}{\bibfnamefont{F.~X.} \bibnamefont{Castellanos}},
  \bibinfo{author}{\bibfnamefont{O.}~\bibnamefont{Sporns}}, \bibnamefont{and}
  \bibinfo{author}{\bibfnamefont{M.~P.} \bibnamefont{Milham}},
  \bibinfo{journal}{Cerebral cortex} \textbf{\bibinfo{volume}{22}},
  \bibinfo{pages}{1862} (\bibinfo{year}{2012}).

\bibitem[{\citenamefont{Sasai et~al.}(2014)\citenamefont{Sasai, Homae,
  Watanabe, Sasaki, Tanabe, Sadato, and Taga}}]{sasai2014frequency}
\bibinfo{author}{\bibfnamefont{S.}~\bibnamefont{Sasai}},
  \bibinfo{author}{\bibfnamefont{F.}~\bibnamefont{Homae}},
  \bibinfo{author}{\bibfnamefont{H.}~\bibnamefont{Watanabe}},
  \bibinfo{author}{\bibfnamefont{A.}~\bibnamefont{Sasaki}},
  \bibinfo{author}{\bibfnamefont{H.}~\bibnamefont{Tanabe}},
  \bibinfo{author}{\bibfnamefont{N.}~\bibnamefont{Sadato}}, \bibnamefont{and}
  \bibinfo{author}{\bibfnamefont{G.}~\bibnamefont{Taga}},
  \bibinfo{journal}{Frontiers in human neuroscience}
  \textbf{\bibinfo{volume}{8}}, \bibinfo{pages}{1022} (\bibinfo{year}{2014}).

\bibitem[{\citenamefont{Thompson and Fransson}(2015)}]{thompson2015frequency}
\bibinfo{author}{\bibfnamefont{W.~H.} \bibnamefont{Thompson}} \bibnamefont{and}
  \bibinfo{author}{\bibfnamefont{P.}~\bibnamefont{Fransson}},
  \bibinfo{journal}{NeuroImage} \textbf{\bibinfo{volume}{121}},
  \bibinfo{pages}{227} (\bibinfo{year}{2015}).

\bibitem[{\citenamefont{Bassett et~al.}(2006)\citenamefont{Bassett,
  Meyer-Lindenberg, Achard, Duke, and Bullmore}}]{bassett2006adaptive}
\bibinfo{author}{\bibfnamefont{D.~S.} \bibnamefont{Bassett}},
  \bibinfo{author}{\bibfnamefont{A.}~\bibnamefont{Meyer-Lindenberg}},
  \bibinfo{author}{\bibfnamefont{S.}~\bibnamefont{Achard}},
  \bibinfo{author}{\bibfnamefont{T.}~\bibnamefont{Duke}}, \bibnamefont{and}
  \bibinfo{author}{\bibfnamefont{E.}~\bibnamefont{Bullmore}},
  \bibinfo{journal}{PNAS} \textbf{\bibinfo{volume}{103}},
  \bibinfo{pages}{19518} (\bibinfo{year}{2006}).

\bibitem[{\citenamefont{Mantini et~al.}(2007)\citenamefont{Mantini, Perrucci,
  Del~Gratta, Romani, and Corbetta}}]{mantini2007electrophysiological}
\bibinfo{author}{\bibfnamefont{D.}~\bibnamefont{Mantini}},
  \bibinfo{author}{\bibfnamefont{M.~G.} \bibnamefont{Perrucci}},
  \bibinfo{author}{\bibfnamefont{C.}~\bibnamefont{Del~Gratta}},
  \bibinfo{author}{\bibfnamefont{G.~L.} \bibnamefont{Romani}},
  \bibnamefont{and} \bibinfo{author}{\bibfnamefont{M.}~\bibnamefont{Corbetta}},
  \bibinfo{journal}{Proceedings of the National Academy of Sciences}
  \textbf{\bibinfo{volume}{104}}, \bibinfo{pages}{13170}
  (\bibinfo{year}{2007}).

\bibitem[{\citenamefont{Supekar et~al.}(2008)\citenamefont{Supekar, Menon,
  Rubin, Musen, Greicius et~al.}}]{supekar2008network}
\bibinfo{author}{\bibfnamefont{K.}~\bibnamefont{Supekar}},
  \bibinfo{author}{\bibfnamefont{V.}~\bibnamefont{Menon}},
  \bibinfo{author}{\bibfnamefont{D.}~\bibnamefont{Rubin}},
  \bibinfo{author}{\bibfnamefont{M.}~\bibnamefont{Musen}},
  \bibinfo{author}{\bibfnamefont{M.~D.} \bibnamefont{Greicius}},
  \bibnamefont{et~al.}, \bibinfo{journal}{PLoS Comput Biol}
  \textbf{\bibinfo{volume}{4}}, \bibinfo{pages}{e1000100}
  (\bibinfo{year}{2008}).

\bibitem[{\citenamefont{Chavez et~al.}(2010)\citenamefont{Chavez, Valencia,
  Navarro, Latora, and Martinerie}}]{chavez2010functional}
\bibinfo{author}{\bibfnamefont{M.}~\bibnamefont{Chavez}},
  \bibinfo{author}{\bibfnamefont{M.}~\bibnamefont{Valencia}},
  \bibinfo{author}{\bibfnamefont{V.}~\bibnamefont{Navarro}},
  \bibinfo{author}{\bibfnamefont{V.}~\bibnamefont{Latora}}, \bibnamefont{and}
  \bibinfo{author}{\bibfnamefont{J.}~\bibnamefont{Martinerie}},
  \bibinfo{journal}{Physical Review Letters} \textbf{\bibinfo{volume}{104}},
  \bibinfo{pages}{118701} (\bibinfo{year}{2010}).

\bibitem[{\citenamefont{Liao et~al.}(2013)\citenamefont{Liao, Xia, Xu, Dai,
  Cao, Niu, Zuo, Zang, and He}}]{liao2013functional}
\bibinfo{author}{\bibfnamefont{X.-H.} \bibnamefont{Liao}},
  \bibinfo{author}{\bibfnamefont{M.-R.} \bibnamefont{Xia}},
  \bibinfo{author}{\bibfnamefont{T.}~\bibnamefont{Xu}},
  \bibinfo{author}{\bibfnamefont{Z.-J.} \bibnamefont{Dai}},
  \bibinfo{author}{\bibfnamefont{X.-Y.} \bibnamefont{Cao}},
  \bibinfo{author}{\bibfnamefont{H.-J.} \bibnamefont{Niu}},
  \bibinfo{author}{\bibfnamefont{X.-N.} \bibnamefont{Zuo}},
  \bibinfo{author}{\bibfnamefont{Y.-F.} \bibnamefont{Zang}}, \bibnamefont{and}
  \bibinfo{author}{\bibfnamefont{Y.}~\bibnamefont{He}},
  \bibinfo{journal}{Neuroimage} \textbf{\bibinfo{volume}{83}},
  \bibinfo{pages}{969} (\bibinfo{year}{2013}).

\bibitem[{\citenamefont{Chen and Glover}(2015)}]{chen2015bold}
\bibinfo{author}{\bibfnamefont{J.~E.} \bibnamefont{Chen}} \bibnamefont{and}
  \bibinfo{author}{\bibfnamefont{G.~H.} \bibnamefont{Glover}},
  \bibinfo{journal}{NeuroImage} \textbf{\bibinfo{volume}{107}},
  \bibinfo{pages}{207} (\bibinfo{year}{2015}).

\bibitem[{\citenamefont{Mucha et~al.}(2010)\citenamefont{Mucha, Richardson,
  Macon, Porter, and Onnela}}]{mucha2010community}
\bibinfo{author}{\bibfnamefont{P.~J.} \bibnamefont{Mucha}},
  \bibinfo{author}{\bibfnamefont{T.}~\bibnamefont{Richardson}},
  \bibinfo{author}{\bibfnamefont{K.}~\bibnamefont{Macon}},
  \bibinfo{author}{\bibfnamefont{M.~A.} \bibnamefont{Porter}},
  \bibnamefont{and} \bibinfo{author}{\bibfnamefont{J.-P.}
  \bibnamefont{Onnela}}, \bibinfo{journal}{Science}
  \textbf{\bibinfo{volume}{328}}, \bibinfo{pages}{876} (\bibinfo{year}{2010}).

\bibitem[{\citenamefont{De~Domenico et~al.}(2013)\citenamefont{De~Domenico,
  Sol{\`e}-Ribalta, Cozzo, Kivel{\"a}, Moreno, Porter, G{\`o}mez, and
  Arenas}}]{dedomenico2013mathematical}
\bibinfo{author}{\bibfnamefont{M.}~\bibnamefont{De~Domenico}},
  \bibinfo{author}{\bibfnamefont{A.}~\bibnamefont{Sol{\`e}-Ribalta}},
  \bibinfo{author}{\bibfnamefont{E.}~\bibnamefont{Cozzo}},
  \bibinfo{author}{\bibfnamefont{M.}~\bibnamefont{Kivel{\"a}}},
  \bibinfo{author}{\bibfnamefont{Y.}~\bibnamefont{Moreno}},
  \bibinfo{author}{\bibfnamefont{M.~A.} \bibnamefont{Porter}},
  \bibinfo{author}{\bibfnamefont{S.}~\bibnamefont{G{\`o}mez}},
  \bibnamefont{and} \bibinfo{author}{\bibfnamefont{A.}~\bibnamefont{Arenas}},
  \bibinfo{journal}{Phys. Rev. X} \textbf{\bibinfo{volume}{3}},
  \bibinfo{pages}{041022} (\bibinfo{year}{2013}).

\bibitem[{\citenamefont{De~Domenico
  et~al.}(2015{\natexlab{a}})\citenamefont{De~Domenico, Nicosia, Arenas, and
  Latora}}]{dedomenico2015structural}
\bibinfo{author}{\bibfnamefont{M.}~\bibnamefont{De~Domenico}},
  \bibinfo{author}{\bibfnamefont{V.}~\bibnamefont{Nicosia}},
  \bibinfo{author}{\bibfnamefont{A.}~\bibnamefont{Arenas}}, \bibnamefont{and}
  \bibinfo{author}{\bibfnamefont{V.}~\bibnamefont{Latora}},
  \bibinfo{journal}{Nature communications} \textbf{\bibinfo{volume}{6}},
  \bibinfo{pages}{6864} (\bibinfo{year}{2015}{\natexlab{a}}).

\bibitem[{\citenamefont{De~Domenico
  et~al.}(2015{\natexlab{b}})\citenamefont{De~Domenico, Sol{\'e}-Ribalta,
  Omodei, G{\'o}mez, and Arenas}}]{dedomenico2015ranking}
\bibinfo{author}{\bibfnamefont{M.}~\bibnamefont{De~Domenico}},
  \bibinfo{author}{\bibfnamefont{A.}~\bibnamefont{Sol{\'e}-Ribalta}},
  \bibinfo{author}{\bibfnamefont{E.}~\bibnamefont{Omodei}},
  \bibinfo{author}{\bibfnamefont{S.}~\bibnamefont{G{\'o}mez}},
  \bibnamefont{and} \bibinfo{author}{\bibfnamefont{A.}~\bibnamefont{Arenas}},
  \bibinfo{journal}{Nature Communications} \textbf{\bibinfo{volume}{6}},
  \bibinfo{pages}{6868} (\bibinfo{year}{2015}{\natexlab{b}}).

\bibitem[{\citenamefont{Kivel{\"a} et~al.}(2014)\citenamefont{Kivel{\"a},
  Arenas, Barthelemy, Gleeson, Moreno, and Porter}}]{kivela2014multilayer}
\bibinfo{author}{\bibfnamefont{M.}~\bibnamefont{Kivel{\"a}}},
  \bibinfo{author}{\bibfnamefont{A.}~\bibnamefont{Arenas}},
  \bibinfo{author}{\bibfnamefont{M.}~\bibnamefont{Barthelemy}},
  \bibinfo{author}{\bibfnamefont{J.~P.} \bibnamefont{Gleeson}},
  \bibinfo{author}{\bibfnamefont{Y.}~\bibnamefont{Moreno}}, \bibnamefont{and}
  \bibinfo{author}{\bibfnamefont{M.~A.} \bibnamefont{Porter}},
  \bibinfo{journal}{Journal of Complex Networks} \textbf{\bibinfo{volume}{2}},
  \bibinfo{pages}{203} (\bibinfo{year}{2014}).

\bibitem[{\citenamefont{Boccaletti et~al.}(2014)\citenamefont{Boccaletti,
  Bianconi, Criado, Del~Genio, G{\'o}mez-Garde{\~n}es, Romance,
  Sendi{\~n}a-Nadal, Wang, and Zanin}}]{boccaletti2014structure}
\bibinfo{author}{\bibfnamefont{S.}~\bibnamefont{Boccaletti}},
  \bibinfo{author}{\bibfnamefont{G.}~\bibnamefont{Bianconi}},
  \bibinfo{author}{\bibfnamefont{R.}~\bibnamefont{Criado}},
  \bibinfo{author}{\bibfnamefont{C.}~\bibnamefont{Del~Genio}},
  \bibinfo{author}{\bibfnamefont{J.}~\bibnamefont{G{\'o}mez-Garde{\~n}es}},
  \bibinfo{author}{\bibfnamefont{M.}~\bibnamefont{Romance}},
  \bibinfo{author}{\bibfnamefont{I.}~\bibnamefont{Sendi{\~n}a-Nadal}},
  \bibinfo{author}{\bibfnamefont{Z.}~\bibnamefont{Wang}}, \bibnamefont{and}
  \bibinfo{author}{\bibfnamefont{M.}~\bibnamefont{Zanin}},
  \bibinfo{journal}{Physics Reports} \textbf{\bibinfo{volume}{544}},
  \bibinfo{pages}{1} (\bibinfo{year}{2014}).

\bibitem[{\citenamefont{Power et~al.}(2011)\citenamefont{Power, Cohen, Nelson,
  Wig, Barnes, Church, Vogel, Laumann, Miezin, Schlaggar
  et~al.}}]{power2011functional}
\bibinfo{author}{\bibfnamefont{J.~D.} \bibnamefont{Power}},
  \bibinfo{author}{\bibfnamefont{A.~L.} \bibnamefont{Cohen}},
  \bibinfo{author}{\bibfnamefont{S.~M.} \bibnamefont{Nelson}},
  \bibinfo{author}{\bibfnamefont{G.~S.} \bibnamefont{Wig}},
  \bibinfo{author}{\bibfnamefont{K.~A.} \bibnamefont{Barnes}},
  \bibinfo{author}{\bibfnamefont{J.~A.} \bibnamefont{Church}},
  \bibinfo{author}{\bibfnamefont{A.~C.} \bibnamefont{Vogel}},
  \bibinfo{author}{\bibfnamefont{T.~O.} \bibnamefont{Laumann}},
  \bibinfo{author}{\bibfnamefont{F.~M.} \bibnamefont{Miezin}},
  \bibinfo{author}{\bibfnamefont{B.~L.} \bibnamefont{Schlaggar}},
  \bibnamefont{et~al.}, \bibinfo{journal}{Neuron}
  \textbf{\bibinfo{volume}{72}}, \bibinfo{pages}{665} (\bibinfo{year}{2011}).

\bibitem[{\citenamefont{van~den Heuvel and Sporns}(2013)}]{van2013network}
\bibinfo{author}{\bibfnamefont{M.~P.} \bibnamefont{van~den Heuvel}}
  \bibnamefont{and} \bibinfo{author}{\bibfnamefont{O.}~\bibnamefont{Sporns}},
  \bibinfo{journal}{Trends in cognitive sciences}
  \textbf{\bibinfo{volume}{17}}, \bibinfo{pages}{683} (\bibinfo{year}{2013}).

\bibitem[{\citenamefont{Rubinov et~al.}(2013)}]{rubinov2013schizophrenia}
\bibinfo{author}{\bibfnamefont{M.}~\bibnamefont{Rubinov}} \bibnamefont{et~al.},
  \bibinfo{journal}{Dialogues in clinical neuroscience}
  \textbf{\bibinfo{volume}{15}}, \bibinfo{pages}{339} (\bibinfo{year}{2013}).

\bibitem[{\citenamefont{Brin and Page}(1998)}]{brin1998}
\bibinfo{author}{\bibfnamefont{S.}~\bibnamefont{Brin}} \bibnamefont{and}
  \bibinfo{author}{\bibfnamefont{L.}~\bibnamefont{Page}}, in
  \emph{\bibinfo{booktitle}{Proceedings of the Seventh International Conference
  on World Wide Web 7}} (\bibinfo{publisher}{Elsevier Science Publishers B.
  V.}, \bibinfo{address}{Amsterdam, The Netherlands, The Netherlands},
  \bibinfo{year}{1998}), WWW7, pp. \bibinfo{pages}{107--117}.

\bibitem[{\citenamefont{Ermann et~al.}(2015)\citenamefont{Ermann, Frahm, and
  Shepelyansky}}]{erman2015google}
\bibinfo{author}{\bibfnamefont{L.}~\bibnamefont{Ermann}},
  \bibinfo{author}{\bibfnamefont{K.~M.} \bibnamefont{Frahm}}, \bibnamefont{and}
  \bibinfo{author}{\bibfnamefont{D.~L.} \bibnamefont{Shepelyansky}},
  \bibinfo{journal}{Rev. Mod. Phys.} \textbf{\bibinfo{volume}{87}},
  \bibinfo{pages}{1261} (\bibinfo{year}{2015}).

\bibitem[{\citenamefont{Breiman}(2001)}]{breiman2001random}
\bibinfo{author}{\bibfnamefont{L.}~\bibnamefont{Breiman}},
  \bibinfo{journal}{Machine learning} \textbf{\bibinfo{volume}{45}},
  \bibinfo{pages}{5} (\bibinfo{year}{2001}).

\bibitem[{\citenamefont{Biswal et~al.}(1995)\citenamefont{Biswal, Yetkin,
  Haughton, and Hyde}}]{biswal1995functional}
\bibinfo{author}{\bibfnamefont{B.}~\bibnamefont{Biswal}},
  \bibinfo{author}{\bibfnamefont{F.~Z.} \bibnamefont{Yetkin}},
  \bibinfo{author}{\bibfnamefont{V.~M.} \bibnamefont{Haughton}},
  \bibnamefont{and} \bibinfo{author}{\bibfnamefont{J.~S.} \bibnamefont{Hyde}},
  \bibinfo{journal}{Magnetic resonance in medicine}
  \textbf{\bibinfo{volume}{34}}, \bibinfo{pages}{537} (\bibinfo{year}{1995}).

\bibitem[{\citenamefont{He}(2011)}]{he2011scale}
\bibinfo{author}{\bibfnamefont{B.~J.} \bibnamefont{He}}, \bibinfo{journal}{The
  Journal of neuroscience} \textbf{\bibinfo{volume}{31}},
  \bibinfo{pages}{13786} (\bibinfo{year}{2011}).

\bibitem[{\citenamefont{van~den Heuvel et~al.}(2013)\citenamefont{van~den
  Heuvel, Sporns, Collin, Scheewe, Mandl, Cahn, Go{\~n}i, Pol, and
  Kahn}}]{van2013abnormal}
\bibinfo{author}{\bibfnamefont{M.~P.} \bibnamefont{van~den Heuvel}},
  \bibinfo{author}{\bibfnamefont{O.}~\bibnamefont{Sporns}},
  \bibinfo{author}{\bibfnamefont{G.}~\bibnamefont{Collin}},
  \bibinfo{author}{\bibfnamefont{T.}~\bibnamefont{Scheewe}},
  \bibinfo{author}{\bibfnamefont{R.~C.} \bibnamefont{Mandl}},
  \bibinfo{author}{\bibfnamefont{W.}~\bibnamefont{Cahn}},
  \bibinfo{author}{\bibfnamefont{J.}~\bibnamefont{Go{\~n}i}},
  \bibinfo{author}{\bibfnamefont{H.~E.~H.} \bibnamefont{Pol}},
  \bibnamefont{and} \bibinfo{author}{\bibfnamefont{R.~S.} \bibnamefont{Kahn}},
  \bibinfo{journal}{JAMA psychiatry} \textbf{\bibinfo{volume}{70}},
  \bibinfo{pages}{783} (\bibinfo{year}{2013}).

\bibitem[{\citenamefont{Glahn et~al.}(2008)\citenamefont{Glahn, Laird,
  Ellison-Wright, Thelen, Robinson, Lancaster, Bullmore, and
  Fox}}]{glahn2008meta}
\bibinfo{author}{\bibfnamefont{D.~C.} \bibnamefont{Glahn}},
  \bibinfo{author}{\bibfnamefont{A.~R.} \bibnamefont{Laird}},
  \bibinfo{author}{\bibfnamefont{I.}~\bibnamefont{Ellison-Wright}},
  \bibinfo{author}{\bibfnamefont{S.~M.} \bibnamefont{Thelen}},
  \bibinfo{author}{\bibfnamefont{J.~L.} \bibnamefont{Robinson}},
  \bibinfo{author}{\bibfnamefont{J.~L.} \bibnamefont{Lancaster}},
  \bibinfo{author}{\bibfnamefont{E.}~\bibnamefont{Bullmore}}, \bibnamefont{and}
  \bibinfo{author}{\bibfnamefont{P.~T.} \bibnamefont{Fox}},
  \bibinfo{journal}{Biological psychiatry} \textbf{\bibinfo{volume}{64}},
  \bibinfo{pages}{774} (\bibinfo{year}{2008}).

\bibitem[{\citenamefont{Ellison-Wright and Bullmore}(2009)}]{ellison2009meta}
\bibinfo{author}{\bibfnamefont{I.}~\bibnamefont{Ellison-Wright}}
  \bibnamefont{and} \bibinfo{author}{\bibfnamefont{E.}~\bibnamefont{Bullmore}},
  \bibinfo{journal}{Schizophrenia research} \textbf{\bibinfo{volume}{108}},
  \bibinfo{pages}{3} (\bibinfo{year}{2009}).

\bibitem[{\citenamefont{van~den Heuvel and Fornito}(2014)}]{van2014brain}
\bibinfo{author}{\bibfnamefont{M.~P.} \bibnamefont{van~den Heuvel}}
  \bibnamefont{and} \bibinfo{author}{\bibfnamefont{A.}~\bibnamefont{Fornito}},
  \bibinfo{journal}{Neuropsychology review} \textbf{\bibinfo{volume}{24}},
  \bibinfo{pages}{32} (\bibinfo{year}{2014}).

\bibitem[{\citenamefont{Yang et~al.}(2010)\citenamefont{Yang, Liu, Sui,
  Pearlson, and Calhoun}}]{yang2010hybrid}
\bibinfo{author}{\bibfnamefont{H.}~\bibnamefont{Yang}},
  \bibinfo{author}{\bibfnamefont{J.}~\bibnamefont{Liu}},
  \bibinfo{author}{\bibfnamefont{J.}~\bibnamefont{Sui}},
  \bibinfo{author}{\bibfnamefont{G.}~\bibnamefont{Pearlson}}, \bibnamefont{and}
  \bibinfo{author}{\bibfnamefont{V.~D.} \bibnamefont{Calhoun}},
  \bibinfo{journal}{Frontiers in human neuroscience}
  \textbf{\bibinfo{volume}{4}} (\bibinfo{year}{2010}).

\bibitem[{\citenamefont{Chyzhyk et~al.}(2015)\citenamefont{Chyzhyk, Savio, and
  Gra{\~n}a}}]{chyzhyk2015computer}
\bibinfo{author}{\bibfnamefont{D.}~\bibnamefont{Chyzhyk}},
  \bibinfo{author}{\bibfnamefont{A.}~\bibnamefont{Savio}}, \bibnamefont{and}
  \bibinfo{author}{\bibfnamefont{M.}~\bibnamefont{Gra{\~n}a}},
  \bibinfo{journal}{Neural Networks} \textbf{\bibinfo{volume}{68}},
  \bibinfo{pages}{23} (\bibinfo{year}{2015}).

\bibitem[{\citenamefont{Honea et~al.}(2005)\citenamefont{Honea, Crow,
  Passingham, and Mackay}}]{honea2005regional}
\bibinfo{author}{\bibfnamefont{R.}~\bibnamefont{Honea}},
  \bibinfo{author}{\bibfnamefont{T.~J.} \bibnamefont{Crow}},
  \bibinfo{author}{\bibfnamefont{D.}~\bibnamefont{Passingham}},
  \bibnamefont{and} \bibinfo{author}{\bibfnamefont{C.~E.}
  \bibnamefont{Mackay}}, \bibinfo{journal}{American Journal of Psychiatry}
  \textbf{\bibinfo{volume}{162}}, \bibinfo{pages}{2233} (\bibinfo{year}{2005}).

\bibitem[{\citenamefont{Anderson et~al.}(2011)\citenamefont{Anderson, Druzgal,
  Lopez-Larson, Jeong, Desai, and Yurgelun-Todd}}]{anderson2011network}
\bibinfo{author}{\bibfnamefont{J.~S.} \bibnamefont{Anderson}},
  \bibinfo{author}{\bibfnamefont{T.~J.} \bibnamefont{Druzgal}},
  \bibinfo{author}{\bibfnamefont{M.}~\bibnamefont{Lopez-Larson}},
  \bibinfo{author}{\bibfnamefont{E.-K.} \bibnamefont{Jeong}},
  \bibinfo{author}{\bibfnamefont{K.}~\bibnamefont{Desai}}, \bibnamefont{and}
  \bibinfo{author}{\bibfnamefont{D.}~\bibnamefont{Yurgelun-Todd}},
  \bibinfo{journal}{Human brain mapping} \textbf{\bibinfo{volume}{32}},
  \bibinfo{pages}{919} (\bibinfo{year}{2011}).

\bibitem[{\citenamefont{De~Domenico
  et~al.}(2015{\natexlab{c}})\citenamefont{De~Domenico, Lancichinetti, Arenas,
  and Rosvall}}]{dedomenico2015identifying}
\bibinfo{author}{\bibfnamefont{M.}~\bibnamefont{De~Domenico}},
  \bibinfo{author}{\bibfnamefont{A.}~\bibnamefont{Lancichinetti}},
  \bibinfo{author}{\bibfnamefont{A.}~\bibnamefont{Arenas}}, \bibnamefont{and}
  \bibinfo{author}{\bibfnamefont{M.}~\bibnamefont{Rosvall}},
  \bibinfo{journal}{Physical Review X} \textbf{\bibinfo{volume}{5}},
  \bibinfo{pages}{011027} (\bibinfo{year}{2015}{\natexlab{c}}).

\bibitem[{\citenamefont{Gomez et~al.}(2013)\citenamefont{Gomez, Diaz-Guilera,
  Gomez-Garde{\~n}es, Perez-Vicente, Moreno, and Arenas}}]{gomez2013diffusion}
\bibinfo{author}{\bibfnamefont{S.}~\bibnamefont{Gomez}},
  \bibinfo{author}{\bibfnamefont{A.}~\bibnamefont{Diaz-Guilera}},
  \bibinfo{author}{\bibfnamefont{J.}~\bibnamefont{Gomez-Garde{\~n}es}},
  \bibinfo{author}{\bibfnamefont{C.~J.} \bibnamefont{Perez-Vicente}},
  \bibinfo{author}{\bibfnamefont{Y.}~\bibnamefont{Moreno}}, \bibnamefont{and}
  \bibinfo{author}{\bibfnamefont{A.}~\bibnamefont{Arenas}},
  \bibinfo{journal}{Physical review letters} \textbf{\bibinfo{volume}{110}},
  \bibinfo{pages}{028701} (\bibinfo{year}{2013}).

\bibitem[{\citenamefont{De~Domenico et~al.}(2014)\citenamefont{De~Domenico,
  Sol\'e-Ribalta, G\'omez, and Arenas}}]{dedomenico2014navigability}
\bibinfo{author}{\bibfnamefont{M.}~\bibnamefont{De~Domenico}},
  \bibinfo{author}{\bibfnamefont{A.}~\bibnamefont{Sol\'e-Ribalta}},
  \bibinfo{author}{\bibfnamefont{S.}~\bibnamefont{G\'omez}}, \bibnamefont{and}
  \bibinfo{author}{\bibfnamefont{A.}~\bibnamefont{Arenas}},
  \bibinfo{journal}{PNAS} \textbf{\bibinfo{volume}{111}}, \bibinfo{pages}{8351}
  (\bibinfo{year}{2014}).

\bibitem[{\citenamefont{Braunstein et~al.}(2006)\citenamefont{Braunstein,
  Ghosh, and Severini}}]{braunstein2006laplacian}
\bibinfo{author}{\bibfnamefont{S.~L.} \bibnamefont{Braunstein}},
  \bibinfo{author}{\bibfnamefont{S.}~\bibnamefont{Ghosh}}, \bibnamefont{and}
  \bibinfo{author}{\bibfnamefont{S.}~\bibnamefont{Severini}},
  \bibinfo{journal}{Annals of Combinatorics} \textbf{\bibinfo{volume}{10}},
  \bibinfo{pages}{291} (\bibinfo{year}{2006}).

\bibitem[{\citenamefont{Passerini and
  Severini}(2010)}]{passerini2010quantifying}
\bibinfo{author}{\bibfnamefont{F.}~\bibnamefont{Passerini}} \bibnamefont{and}
  \bibinfo{author}{\bibfnamefont{S.}~\bibnamefont{Severini}},
  \bibinfo{journal}{Developments in Intelligent Agent Technologies and
  Multi-Agent Systems: Concepts and Applications: Concepts and Applications}
  p.~\bibinfo{pages}{66} (\bibinfo{year}{2010}).

\end{thebibliography}

\end{document}